\documentclass[prb,twocolumn,10pt,aps,longbibliography,superscriptaddress]{revtex4-2}

\usepackage{bm}   %Boldsymbol in math mode
\usepackage{amsmath}
\usepackage{amssymb}
\usepackage{amsfonts}
\usepackage{mathrsfs}
\usepackage{comment}
\usepackage{bm}

\usepackage{graphicx}
\usepackage{color}
\usepackage{dcolumn}  % Aligns table columns on decimal point
\usepackage{soul}     % Provides hyphenatable spacing out (letterspacing), underlining, striking out, etc.
\usepackage{multirow}
\usepackage{hyperref}
\usepackage{cancel}

\usepackage{comment} %Selectively include/exclude pieces of text
\usepackage{epstopdf} %Converts EPS file to PDF

\newcommand{\tcol}[1]{\textcolor{black}{#1}}
\newcommand{\newcol}[1]{\textcolor{black}{#1}}

\newcommand{\Ha}{\mathcal{H}}
\newcommand{\Z}{\mathcal{Z}}
\newcommand{\de}{\text{d}}
\newcommand{\D}{\mathcal{D}}
\newcommand{\bs}{\boldsymbol{\sigma}}

\begin{document}
%%Title, authors, affiliations, abstract
\title{Probing marginal stability in the spherical $p=2$ model}

\author{Jacopo Niedda}
\affiliation{ICTP, Strada Costiera 11, 34151, Trieste, Italy}

\author{Tommaso Tonolo}
\affiliation{Gran Sasso Science Institute, Viale F. Crispi 7, 67100 L’Aquila, Italy}
\affiliation{INFN-Laboratori Nazionali del Gran Sasso, Via G. Acitelli 22, 67100 Assergi (AQ), Italy}

\author{Giacomo Gradenigo}
\affiliation{Gran Sasso Science Institute, Viale F. Crispi 7, 67100 L’Aquila, Italy}
\affiliation{INFN-Laboratori Nazionali del Gran Sasso, Via G. Acitelli 22, 67100 Assergi (AQ), Italy}

\begin{abstract}
  In this paper we investigate the marginally stable nature of the
  low-temperature trivial spin glass phase in the spherical $p=2$ spin
  glass, by perturbing the system with three different kinds of
  non-linear interactions. In particular, we compare the effect of
  three additional dense four-body interactions: ferromagnetic
  couplings, purely disordered couplings and couplings with competing
  disordered and ferromagnetic interactions. Our study, characterized
  by the effort to present in a clear and pedagogical way the
  derivation of all the results, shows that the marginal stability
  property of the spherical spin glass depends in fact on which kind
  of {\it perturbation} is applied to the system: in general, a
  certain degree of frustration is needed also in the additional terms
  in order to induce a transition from a trivial to a non-trivial
  spin-glass phase. On the contrary, the addition of generic
  non-frustrated interactions does not destabilize the trivial
  spin-glass phase.
  
\end{abstract}

\maketitle
%\tableofcontent

\section{Introduction}
Marginal stability is one among the most interesting and fascinating
features of multi-agent systems with frustrated interactions,
something that enjoys at the same time a precise mathematical
description and encodes properties which meet the physical intuition
and can be explained with clear words. First of all, marginal
stability is something that makes really sense only for systems with a
large number of degrees of freedom, $N\gg 1$: it corresponds to the
existence of stable configurations which may change macroscopically
under the effect of perturbing just few variables. More precisely a
number of variables which do not scale with the size of the system. It
can be immediately realized that this is a very peculiar
behaviour. What usually characterizes the behaviour of physical
systems and is also in agreement with the intuition that we have from
ordinary life, is that to produce extensive changes in a system we
have to exploit an extensive perturbation. This is also the physical
content of the linear-response theorem~\cite{K57}. Just to make an
example, it is very natural to expect that in order to bend a plastic
stick we need to exploit an effort proportional to the effect we wish
to obtain.

At the same time we know that also {\it catastrophes} take
place. Consider for instance the stability of an ecosystem with a very
large number of species, a subject which has recently attracted a lot
of interest from the community working on disordered
systems~\cite{Biroli18,Altieri19,Altieri21}. A catastrophe is for
instance represented by the sudden extinction of a particular species,
due to environmental changes, which then triggers a mass extinction of
many other species in a cascade and the appearance of new ones. This
is an example of macroscopic change produced by a small
perturbation. Another example from the recent literature on disordered
systems is the response to external perturbations in the jamming phase
of hard spheres~\cite{CKPUZ14b,CKPUZ17,PaUrZa20,CKPUZ14a,KPUZ13}: it
is the marginally stable nature of the low-lying minima in the
free-energy landscape characteristic of continuous replica symmetry
breaking~\cite{Parisi79,Mezard87} that allows to explain the low-cost
excitations responsible of the critical properties close to the
jamming transition. The previous ones are just two examples to grasp
the importance of the marginal stability idea. Technically, in the
context of disordered systems, marginal stability refers to the
situation where stationary points of a thermodynamic potential
corresponding to non-trivial low-temperature glassy phases have a
linear stability matrix with respect to overlap fluctuations with null
eigenvalues, i.e., flat directions where costless excitations may
trigger macroscopic change in the system. These flat directions in the
potential landscape are associated with the divergence in the
spin-glass susceptibility, meaning that the system is extremely
sensible to small perturbations.

As a tribute to the importance played nowadays in many different
contexts of marginal stability, in the present work we wish to present
a detailed and comprehensive analysis of this property in a
paradigmatic model of spin-glasses: the spherical spin glass. It is
well known that in the Sherrington-Kirkpatrick (SK) model \cite{SK75}
there is a transition to a non-trivial spin-glass phase at a certain
temperature $T_c$. Let us recall that the SK model is defined by the
Hamiltonian
\begin{align}
H_J = -\sum_{i<j} J_{ij} S_i S_j,
\label{eq:model-hamiltonian}
\end{align}
where the summation runs over all pairs of independent indices and
$S_i=\lbrace -1,+1 \rbrace$ are Ising spins. The coefficients $J_{ij}$
are random quenched variables with a zero-mean Gaussian distribution
and variance $\sigma^2_J = J^2/N$. In this case, i.e.~in absence of an
external magnetic field, the transition temperature is simply given by
$T_c=J$. In the spin-glass phase the overlap $q = N^{-1}\sum_{i=1}^N
S_i^a S_i^b$ between two equilibrium configurations at a given
temperature, labeled by $a$ and $b$, can take infinitely many values
according to a non-trivial distribution $P(q)$, with $q \in
[0,1]$~\cite{Parisi79}. The correct form of the distribution $P(q)$
was obtained by Parisi, in the so-called continuous Replica Symmetry
Breaking (full-RSB) solution scheme~\cite{Parisi79}, which correctly
describes the low temperature phase of the model. \newcol{An
  interesting work on the marginal stability of the SK model is
  \cite{Muller06}, where the study of marginally stable states is
  tackled with different methods, i.e.~both replica and cavity
  computations, which all reveal the presence of a soft mode in the
  free-energy landscape.}

It was already known before Parisi's solution that if we consider the
same Hamiltonian of Eq.~\eqref{eq:model-hamiltonian} where in place of
the discrete Ising spins one has continuous variables subject to a
spherical constraint, i.e., $\sigma_i\in \mathbb{R}$ with $N =
\sum_{i=1}^N \sigma_i^2$, there is a phase transition at the same
critical temperature $T_c=J$ of the SK model, but in this case the low
temperature phase is a {\it trivial} spin-glass phase. The solution of
the spherical model was obtained by Kosterlitz {\it et al}
\cite{Kosterlitz76} by using random matrix theory. In the language of
replicas, the low temperature phase of the model is still Replica
Symmetric (RS) and is characterized by a trivial probability
distribution of the overlap: a Dirac delta, $P(q) = \delta(q-q_T^*)$,
where $q_T^*$ is a finite equilibrium value which depends on the
temperature.

What is very peculiar of this trivial spin-glass phase, sometimes
called {\it disguised ferromagnet}, is that it is marginally stable,
in the sense of having a vanishing mode of the stability matrix, even
if there is no breaking of the replica symmetry
\cite{DeDominicis06}. One can therefore legitimately wonder with
respect to which kind of perturbations, intended as additional terms
in the Hamiltonian, the system is marginally stable and try to relate
the nature of the perturbation to the new stable phase where the
system is driven by the perturbation. Is it true that the marginally
stable trivial spin glass phase of the spherical model becomes
unstable in favour of a true spin glass phase by any sort of
perturbation of the Hamiltonian? \newcol{An investigation of this
  kind, when the non-linear perturbation is disordered, is the one
  discussed in terms of supersymmetry breaking states
  in~\cite{Annibale04}. Here we are interested in more generic kind of
  perturbations, also including ordered ferromagnetic interactions.}
For instance: do non-linear ordered and disordered additional terms
have similar or different effects?  What kind of phases will these
perturbations induce, when the RS trivial spin-glass phase is
destabilized? The aim of this paper is to provide a clear answer to
these questions, through a detailed and didactic presentation aimed at
the wider possible audience.

The paper is organized as follows: in Sec.~\ref{sec:p2-model} we
introduce the spherical spin-glass model and in
Sec.~\ref{sec:non-linear-ordered} we study the effect of adding an
ordered non-linear interaction term of the kind $(\epsilon/
N^3)\sigma_i\sigma_j\sigma_k\sigma_l$, where a summation over all
independent quadruplets of indices is considered, on the
low-temperature marginally stable phase of the system, where
$\sigma_i\in\mathbb{R}$: the phase diagram in the plane $(\epsilon,T)$
is discussed. In Sec.~\ref{sec:non-linear-disordered} we study the
effect of a purely disordered perturbation, namely we introduce non
linear couplings of the form $ J_{ijkl}
\sigma_i\sigma_j\sigma_k\sigma_l$, where $J_{ijkl}$ are random
coefficients following a Gaussian distribution with {\it zero} mean
and variance proportional to $\epsilon^2/N^3$, which guarantees the
extensivity of the Hamiltonian. The phase diagram in the plane
$(\epsilon,T)$ is presented, paralleling the results of
Refs.~\cite{Crisanti04,Crisanti06}. Finally in
Sec.~\ref{sec:non-linear-mixed} we study the competing effect of
ordered and disordered non-linear interactions, assigned in the form
of non-linear couplings $J_{ijkl} \sigma_i\sigma_j\sigma_k\sigma_l$
where the Gaussian distribution of the random coefficients has both
mean and variance different from zero, controlled by the same
parameter $\epsilon$. Also in this case we discuss the phase diagram
in the $(\epsilon,T)$ plane. The difference with the corresponding
phase diagram in Sec.~\ref{sec:non-linear-ordered} is that now, by
increasing $\epsilon$, the strength of both ordered and disordered
terms in the non-linear interaction is increased at the same time. In
Sec.~\ref{sec:conclusions} some conclusions and perspectives are
drawn. Part of the derivations proposed here goes along the same path
of those for the result already presented in
Ref.~\cite{Crisanti06}. But, in order to make the presentation as
clear as possible, we have repeated them in the context of the present
framework.

\section{The spherical $p=2$ model}
\label{sec:p2-model}

The spherical $p=2$ model \cite{Kosterlitz76} is defined by the
Hamiltonian
\begin{equation} \label{Ham_p=2}
    \Ha_J(\bs) = - \sum_{i<j}^{1,N} J_{ij}~\sigma_i\sigma_j,
\end{equation}
where $\bs \in \mathbb{R}^N$ is an array of $N$, locally unbounded,
real variables subjected to a global spherical constraint:

\begin{align}
\bs \cdot \bs = \sum_{i=1}^N \sigma_i^2 = N.
\end{align}

The couplings $J_{ij}$ are i.i.d. random variables taken from the
Gaussian distribution

\begin{align}
p(J_{ij}) = \frac{1}{\sqrt{2\pi\sigma_2^2}}e^{-\frac{(J_{ij}-J_0)^2}{2\sigma_2^2}},
\label{eq:J2_gauss}
\end{align}
where $\sigma_2^2=J^2/N$ and $J$ is a free parameter. For the pure
spherical model one usually considers $J_0=0$. The scaling of the
variance ensures that the Hamiltonian is extensive in the large-$N$
limit. For any given instance of the random coefficients $J_{ij}$, the
probability distribution of the spin configurations is given by
\begin{align}
P_{\beta,J}(\bs)= \frac{1}{\Z_J(\beta)} e^{-\beta \Ha_J(\bs)}\delta\left(N-\bs\cdot\bs\right),
\end{align}
where the partition function $\Z_J(\beta)$ reads
\begin{equation} \label{1-PartFunc2}
\Z_J(\beta) = \int \left(\prod_{i=1}^N d\sigma_i\right)~e^{-\beta
  \Ha_J(\bs)} \delta\left(N-\bs\cdot\bs\right).
\end{equation}

We know that thanks to self-averaging, in the large-$N$ limit the
typical free energy corresponds to the average over disorder:
\begin{align} \label{1-FreeEn1}
\beta \overline{f_J} = \lim_{N \rightarrow \infty} -\frac{1}{\beta
  N}~\overline{\ln \Z},
\end{align}
where the overline denotes
\begin{align}
\overline{(\cdots)} = \int \prod_{i<j}^{1,N}
dJ_{ij}~p(J_{ij})(\cdots).
\end{align}

\begin{figure}  
    \includegraphics[width=\columnwidth]{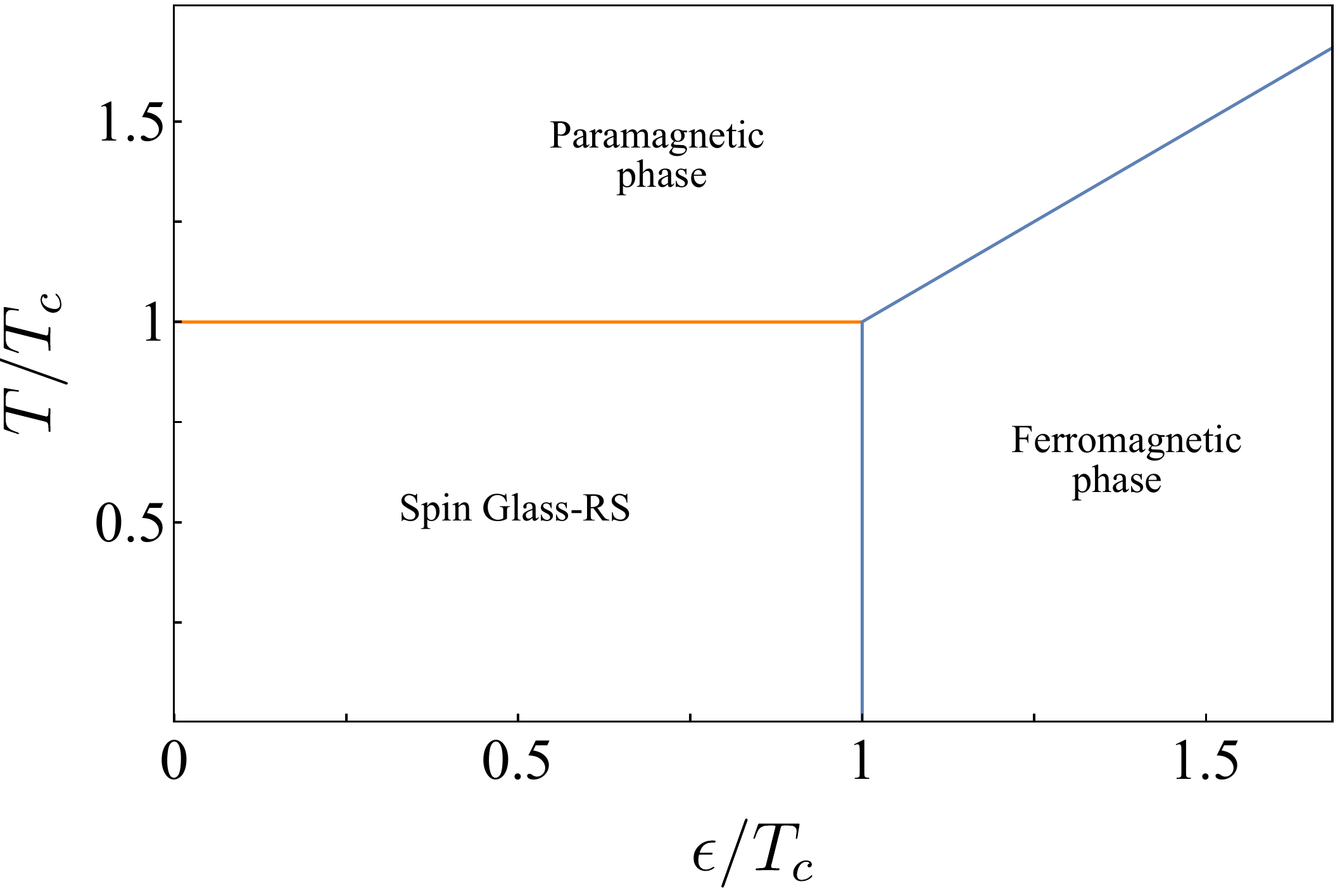}
    \caption{Phase diagram in the $(\epsilon,T)$ plane for the $p=2$
      spherical spin glass, where $\epsilon$ represents the strength
      of two-body ferromagnetic interactions.}
    \label{fig:PhaseDiag0}
\end{figure}

The model can be solved either with the replica method or by
exploiting random matrix theory~\cite{Kosterlitz76},
since the $J_{ij}$'s define the elements of a matrix from the
Gaussian Orthogonal Ensemble (GOE). By solving the model with replicas
one finds that the RS ansatz, $q_{ab} =
\delta_{a b} + (1-\delta_{ab}) q$, where
$q_{ab}$ is the matricial order parameter which enters all
replica calculations, is stable at all temperatures, with a transition
from zero to finite average overlap at $T_c=J$. The exact solution of the model shows that different kind of perturbations have very different effects. If a linear coupling with
an external magnetic field is introduced, by considering a total
Hamiltonian of the kind $\mathcal{H}_{\text{tot}}(\bs)=
\mathcal{H}_{h}(\bs)+\mathcal{H}_{J}(\bs)$, where
$\mathcal{H}_{J}(\bs)$ is the one of Eq.~\eqref{Ham_p=2} and
$\mathcal{H}_h(\bs) = h \sum_{i=1}^N\sigma_i$, one finds that the
magnetic field completely washes the transition away. This means that drawing  the phase diagram of the model in the plane $(h,T)$
 the only
critical point is at $(h=0,T=J)$ and for all values $h\neq 0$ there are
no phase transitions. The behaviour is different if we take into
account ferromagnetic couplings in addition to random couplings, i.e.,
if we allow the parameter $J_0$ to be different from zero in the
distribution of Eq.~\eqref{eq:J2_gauss}. In this case one finds that
the low-temperature trivial spin-glass phase of the spherical model is
in fact stable with respect to an increase of $J_0$, until a
transition to a ferromagnetic phase occours at large enough $J_0$. This is shown in Fig.~\ref{fig:PhaseDiag0}, which simply reproduces the phase diagram
found by Kosterlitz {\it et al.} in Ref.~\cite{Kosterlitz76}.

We will now analyze in detail the behaviour of the spherical
$p=2$ spin model with $J_0=0$ (i.e., zero mean of linear couplings
distribution) when different kinds of non-linear terms are added to the
Hamiltonian. In practice we will study different realizations of the
non-linear interactions for the $2+p$-spin model ($p$ is the order of
the non-linear interactions), for which, in the case of purely
disordered interactions, a careful analysis can be also found
in Refs.~\cite{Crisanti04,Crisanti06}. From the general perspective,
physically interesting application can be found in the modelling of
light modes interactions in random
lasers~\cite{Antenucci2015,GVDBGLC15,GAL20,NLG23,NGLP23}.\\

\section{Ordered non-linearity} \label{sec:non-linear-ordered}
The first perturbation to the spherical $p=2$ model that we have studied is
represented by an ordered 4-body interaction term. The model is
defined by the Hamiltonian
\begin{align} \label{1_Ham}
\Ha_{\text{ord}}(\bs) = \Ha_J(\bs) - \frac{4!\epsilon}{N^3}\sum_{i<j<k<l}^{1,N} \sigma_i\sigma_j\sigma_k\sigma_l,
\end{align}
where $\Ha_J$ is the Hamiltonian~\eqref{Ham_p=2} of the spherical
$p=2$ model with $J_0=0$. The scaling of the ordered coupling
magnitude is chosen in order for the non-linear term to be extensive. The
parameter $\epsilon$ can be tuned to probe different regimes
according to the strength of the non-linearity. The free energy of the model is obtained by means of a standard
replica approach, which provides the exact mean-field solution in the
large-$N$ limit.

For this model one has to consider two global order parameters
\begin{align}
m_a &= \frac{1}{N} \sum_{i=1}^N \sigma_i^a \nonumber \\
q_{ab} &= \frac{1}{N} \sum_{i=1}^N \sigma_i^a\sigma_i^b,
\end{align}
where $m_a$ and $q_{ab}$ are respectively the magnetization elements
and the overlap matrix elements. From here on, we will use the
following symbols to denote the full overlap matrix and magnetization
vector in replica space: $\mathbb{Q} = \lbrace
q_{ab}\rbrace_{a,b=1,\ldots,n}$ and ${\bf m} = \lbrace m_a
\rbrace_{a=1,\ldots,n}$. In terms of these order parameters the
replicated partition function averaged over the disorder reads as
\begin{align} \label{1-PartFunc}
\overline{\Z^n} &= \int \prod_{a<b}^{1,n} \de q_{ab}\int \prod_{a=1}^n  \de m_a~e^{-N G[\mathbb{Q},{\bf m}]},
\end{align}
where the action functional $G$ is defined as
  \begin{align}
    \label{1-Action}
G[\mathbb{Q},{\bf m}] = &- \frac{(\beta J)^2}{4} \sum_{ab}^{1,n} q_{ab}^2 - \beta\epsilon  \sum_{a=1}^n m_a^4 \\
        &- \frac{1}{2} \ln\det (\mathbb{Q}-{\bf m} \otimes {\bf m}^T),
\end{align}
with 
\begin{equation}\label{1-freeEnergy}
    \beta \overline{f_J} = \lim_{n\rightarrow 0} \frac{1}{n}~\min_{m_a}~\max_{q_{ab}}~G[q_{ab},m_a].
\end{equation}

In Eq.~\eqref{1-Action} the symbol ${\bf m} \otimes {\bf m}^T$ denotes
a matrix with elements $({\bf m} \otimes {\bf
  m}^T)_{ab}=m_am_b$. Details on the derivation of $G[\mathbb{Q},{\bf
    m}]$ are given in App.~\ref{AppA}. As in standard replica
approach, in order to find the physical value of the free energy we
have to maximize with respect to the matricial order parameters and
minimize with respect to the vectorial ones
\cite{Mezard87}. Stationary points are found by solving the following
saddle-point equations:

\begin{align} 
  \label{1-SPeqs-a}
  \frac{\delta G}{\delta q_{ab}}  &= (\beta J)^2 q_{ab} +  (\mathbb{Q}-{\bf m} \otimes {\bf m}^T)^{-1}_{ab} = 0, \\
  \label{1-SPeqs-b}
  \frac{\delta G}{\delta m_a}  &= - 4\beta\epsilon~ m_a^3 +
       [(\mathbb{Q}-{\bf m} \otimes {\bf m}^T)^{-1} \cdot {\bf m}]_a = 0.
\end{align}

In what follows we will present the results obtained within the RS
ansatz for the matrix $q_{ab}$, which turns out to be stable for all
values of temperature $T$ and strength $\epsilon$ of the non-linear
coupling.

\subsection{Replica Symmetric Ansatz}
Let us then assume for the matrix $\mathbb{Q}$ a RS ansatz
\begin{align}\label{rs-ansatz}
q_{ab} = \delta_{ab} + (1-\delta_{ab})~q_0
\end{align}

For what concerns the vector ${\bf m}$ we must assume that all
elements are identical, since physical quantities cannot depend on the
replicas. By doing so we get

\begin{align}\label{inversematrix}
    \left[(\mathbb{Q}-{\bf m} \otimes {\bf m}^T)^{-1}\right]_{ab} = \left \{ \begin{array}{rl}
\cfrac{1-2q_0 + m^2}{(1-q_0)^2}~~~~a=b  \\[12pt]
-\cfrac{q_0-m^2}{(1-q_0)^2}~~~~a \neq b.
\end{array}
\right.
\end{align}

The RS effective action reads, in the limit $n\rightarrow 0$, as

\begin{equation} \label{1-RSFreeEn}
\begin{aligned}
   \lim_{n\rightarrow 0} \frac{1}{n} G[q_0,m] = &- \frac{(\beta J)^2}{4} (1-q_0^2) - \beta\epsilon m^4  \\
     &- \frac{1}{2}\ln(1-q_0) - \frac{1}{2} \frac{q_0-m^2}{1-q_0}. 
\end{aligned}
\end{equation}

The RS saddle-point equations are obtained by plugging the ansatz
\eqref{rs-ansatz} in Eqs.~\eqref{1-SPeqs-a},\eqref{1-SPeqs-b} or by
deriving Eq.~\eqref{1-RSFreeEn} with respect to $q_0$ and $m$:

\begin{align} 
\frac{\partial G}{\partial q_0} =& (\beta J)^2 q_0 - \frac{q_0 - m^2}{(1-q_0)^2} = 0 \label{1-RSspEqsALLSOL1} \\
\frac{\partial G}{\partial m} =& -4\beta\epsilon m^3 + \frac{m}{1-q_0} = 0. \label{1-RSspEqsALLSOL2}  
%\label{1-RSspEqsALLSOL}
\end{align}

A trivial solution for Eq.~\eqref{1-RSspEqsALLSOL2} is $m=0$: in this
case Eq.~\eqref{1-RSspEqsALLSOL1} simply becomes identical to the RS
saddle-point equation for the original $p=2$ spherical model:

\begin{align}
[1-(\beta J)^2 (1-q_0)^2] q_0 = 0.
\end{align}

The above equation has two solutions, corresponding to the two
possible phases of the model: the paramagnetic phase with $q_0=0$, and
the trivial spin-glass phase with $q_0=1-\frac{1}{\beta
  J}$. Eq.~\eqref{1-RSspEqsALLSOL2} admits also a $m\neq 0$ solution,
which reads:
\begin{align} \label{1-RSspEq1}
m^2 = \frac{1}{4\beta \epsilon} \frac{1}{1-q_0},
\end{align}
that plugged into the equation for $q_0$ gives
\begin{align} \label{1-RSspEq2}
q_0 (1-q_0)^3\left[\frac{1}{1-q_0} - (\beta J)^2 \right] = \frac{1}{4 \beta \epsilon}.
\end{align}

Summarizing, the analysis of the saddle-point equations obtained by
assuming a RS ansatz yields overall three kind of
solutions: the paramagnetic one with both $m=0$ and $q_0=0$; the
solution corresponding to the trivial spin glass with $m=0$ and
$q_0=1-1/(\beta J)$ and a ferromagnetic solution where both $m$ and
$q_0$ are different from zero and are given by the solutions of
Eqs.~\eqref{1-RSspEq1} and \eqref{1-RSspEq2}. The first two solutions
are the only possible ones with $\epsilon=0$, that is the unperturbed
case. The third solution is the interesting one in presence of
non-linearity and can be studied numerically.

The following step of our analysis will consist in studying whether
the {\it ordered} quartic non-linearity may trigger or not an
instability of the RS ansatz, in particular whether or
not it drives, in the low temperature regime, a transition from a
trivial spin-glass phase to a true spin-glass phase with a non-trivial
distribution of the overlap. In order to do that we need to study the
stability of the RS solutions, which is the subject of
the following section.

\subsection{Stability of the RS solution and phase diagram} \label{1-Stability}

The nature of the fluctuations around the saddle point determines the
stability of the solutions. By following \cite{deAlmeida78}, it is convenient to define
a single array containing all variational parameters: 

\begin{align}
\boldsymbol{\eta} = \binom{{\bf m}}{[\mathbb{Q}]_{a\neq b}}.
\label{eq:array-fluct}
\end{align} 

The object defined in Eq.~\eqref{eq:array-fluct} is a vector in an
\tcol{$n(n+1)/2$}-dimensional space, with element $\eta_A$ where the subscript
index takes values in the range \tcol{$A=1,\ldots,n(n+1)/2$}. The number of
variational parameters, with respect to which fluctuations must be
taken into account, is \tcol{$n(n+1)/2$} because we have to consider
the $n$ components of ${\bf m}$ and $n(n-1)/2$ components for
$\mathbb{Q}$, since for the overlap matrix the diagonal elements are
fixed to $q_{aa}=1$ by the spherical constraint. The effective action
\eqref{1-Action}~can be expanded around the saddle point
$\boldsymbol{\eta}^* = (\boldsymbol{m}^*,\mathbb{Q}^*)$ up to the
second order in the deviations from saddle-point solutions:

\begin{align} \label{1-Action2secOrd1}
G[\boldsymbol{\eta}^*+\delta\boldsymbol{\eta} ]=G[\boldsymbol{\eta}^*] + \frac{1}{2} \sum_{AB} \frac{\partial^2 G [\boldsymbol{\eta}^*]}{\partial \eta_A \partial \eta_B}~\delta \eta_A \delta \eta_B + \ldots, \nonumber \\
\end{align}

In Eq.~\eqref{1-Action2secOrd1} there are no linear terms because when
$\boldsymbol{\eta}^*$ is selected as the solution of saddle-point
equations we have by definition

\begin{align}
  \frac{\partial G[\boldsymbol{\eta}^*]}{\partial \eta_A}=0 \quad\quad\quad \forall~A.
\end{align}  
In order to assess the stability of the solution $\boldsymbol{\eta}^*$
one has to study the spectrum of the Hessian
$\mathbb{H}_{AB}(\boldsymbol{\eta}^*) = \partial^2
G[\boldsymbol{\eta}^*] / \partial \eta_A \partial \eta_B$. By
explicitating the dependence on the overlap and magnetization
fluctuations, the expansion in Eq.~\eqref{1-Action2secOrd1} can be
rewritten as:

\begin{widetext}

\begin{equation}  \label{1-Action2secOrd2}
G[\mathbb{Q},{\bf m}] = G[\mathbb{Q}^*,{\bf m}^*] + \frac{1}{2}\sum_{ab} \frac{\partial^2 G[\mathbb{Q}^*,{\bf m}^*]}{\partial m_a\partial m_b} \delta m_a \delta m_b + \sum_{ab,c} \frac{\partial^2 G[\mathbb{Q}^*,{\bf m}^*]}{q_{ab}\partial m_c}  \delta q_{ab} \delta m_c + \frac{1}{2}\sum_{ab,cd}\frac{\partial^2 G[\mathbb{Q}^*,{\bf m}^*]}{\partial q_{ab}\partial q_{cd}} \delta q_{ab} \delta q_{cd}  + ...
\end{equation} 

\end{widetext}

It is well known that, in the limit $n\rightarrow 0$, the smallest
eigenvalue of the Hessian $\mathbb{H}[\boldsymbol{\eta}^*]$ is the
replicon~\cite{deAlmeida78}, which is related only to overlap fluctuations and can be obtained from the
diagonalization of the submatrix
\begin{equation}
    G_{(ab),(cd)} = \frac{\partial^2 G[\mathbb{Q}^*,{\bf m}^*]}{\partial q_{ab}\partial q_{cd}}.
\end{equation}
In general, i.e.~without specifying the solution ansatz for the saddle
point, this matrix has three different kinds of elements, defined by
taking: (\emph{i}) $a=c$ and $b=d$; (\emph{ii}) either $a\neq c$ and
$b=d$ or $a= c$ and $b\neq d$; (\emph{iii}) $a\neq c$ and $b\neq
d$. These elements are related to different correlations of the
replicated local variables, as can be immediately recognized in the
case of a model with Ising variables (see Ref.~\cite{deAlmeida78} for the RS case). In the case
of spherical variables, we can rewrite the action functional
\eqref{1-Action} in the following way

\begin{equation}
    \begin{aligned}
    G[\mathbb{Q}&,{\bf m}] = - \frac{(\beta J)^2}{2} \sum_{a < b}^{1,n} q_{ab}^2 - \beta\epsilon  \sum_{a=1}^n m_a^4 \\
        & + \ln \int \prod_{a=1}^n \de x_a e^{-\sum_{a \leq b}^n[(\mathbb{Q}-{\bf m} \otimes {\bf m}^T)^{-1}]_{ab} x_a x_b },
\end{aligned}
\end{equation}
where we have introduced back the local variables as correlated
Gaussian auxiliary variables to represent the entropic term ~$\ln\det
(\mathbb{Q}-{\bf m} \otimes {\bf m}^T)$.  With this formalism, we have
\begin{equation}
    \frac{\partial G[\mathbb{Q},{\bf m}]}{\partial q_{a b}} = - (\beta J)^2 q_{a b} - \langle x_a x_b \rangle,
\end{equation}
where the average is computed over the Gaussian distribution of the
$x_a$ variables:

\begin{equation}
    \langle \cdots \rangle = \frac{\int \prod_{a=1}^n \de x_a e^{-\sum_{a \leq b}^n(\mathbb{Q}-{\bf m} \otimes {\bf m}^T)_{ab} x_a x_b } (\cdots )}{\int \prod_{a=1}^n \de x_a e^{-\sum_{a \leq b}^n(\mathbb{Q}-{\bf m} \otimes {\bf m}^T)_{ab} x_a x_b }}.
\end{equation}

Therefore, by using the fact that
\begin{equation}
    \frac{\partial }{\partial q_{c d}} \langle x_a x_b \rangle = - \langle x_a x_b x_c x_d \rangle + \langle x_a x_b \rangle\langle x_c x_d \rangle,
\end{equation}
we have 
\begin{subequations} \label{Hessian_elements}
    \begin{align}
        & G_{(ab),(ab)} \equiv \frac{\partial^2 G}{\partial q_{ab}^2} = -(\beta J)^2 + \langle x_a^2 x_b^2 \rangle - \langle x_a x_b \rangle^2 \\
        & G_{(ab),(ac)} \equiv \frac{\partial^2 G}{\partial q_{ac} \partial q_{ab} } = \langle x_a^2 x_b x_c \rangle - \langle x_a x_b \rangle\langle x_a x_c \rangle  \\
        & G_{(ab),(cd)} \equiv \frac{\partial^2 G}{\partial q_{cd} \partial q_{ab} } = \langle x_a x_b x_c x_d \rangle -\langle x_a x_b \rangle\langle x_c x_d \rangle.
    \end{align}
\end{subequations}

We can now use Wick's theorem to compute the contractions of the 4-point Gaussian correlators and express them in terms of the second moment of the Gaussian distribution. We have
  \begin{subequations} \label{1-GHessElem}
    \begin{align}
        &G_{(ab),(ab)} = -(\beta J)^2 + \langle x_a^2 \rangle \langle x_b^2 \rangle + \langle x_a x_b \rangle^2 \\
        &G_{(ab),(ac)} = \langle x_a^2\rangle \langle x_b x_c \rangle + \langle x_a x_b \rangle\langle x_a x_c \rangle  \\
        &G_{(ab),(cd)}  = \langle x_a x_c \rangle \langle x_b x_d \rangle + \langle x_a x_d \rangle\langle x_b x_c \rangle,
    \end{align}
\end{subequations}
where 
\begin{equation}
    \langle x_a x_b \rangle = \left[(\mathbb{Q}-{\bf m} \otimes {\bf m}^T)^{-1}\right]_{ab}.
\end{equation}

\subsubsection{Replicon}

The diagonalization of $G_{(ab),(cd)}$ has to be performed case by
case depending on the solution ansatz for the saddle point problem. If
$(\mathbb{Q}^*,{\bf m}^*)$ is a RS saddle point, due to
the symmetry of the overlap matrix $\mathbb{Q}^*$ under the
permutation of $n$ replicas, $G_{(ab),(cd)}$ depends only on three
numbers:

\begin{subequations}
    \begin{align}
        &P = -(\beta J)^2 + \langle x_a^2 \rangle^2 + \langle x_a x_b \rangle^2 \\
        &Q = \langle x_a^2\rangle \langle x_a x_b \rangle + \langle x_a x_b \rangle^2  \\
        &R  = 2 \langle x_a x_b \rangle^2.
    \end{align}
\end{subequations}

Therefore, the eigenvalues of $G_{(ab),(cd)}$ depend on combinations
of $P,Q$ and $R$ and can be computed in a straightforward manner by
following the procedure of Ref.~\cite{deAlmeida78} or by using the
Replica Fourier Transform \cite{DeDominicis06,Crisanti15}. There are
three types of eigenvalues at finite $n$: the \emph{longitudinal} one,
corresponding to fluctuations of the overlap which do not break the
replica symmetry, the \emph{anomalous} one, corresponding to
fluctuations which break the replica symmetry, but violate the
property of the overlap matrix of having the same sum over each row,
and the \emph{replicon}, connected to fluctuations which break the
replica symmetry preserving the properties of the overlap matrix. In
the limit $n \rightarrow 0$, the longitudinal and anomalous
eigenvalues degenerate in the same one. The replicon has the following
expression:

\begin{equation} \label{1-replicon}
\begin{split}
   \lambda_R &= P - 2Q + R \\
    &= -(\beta J)^2 + \left( \langle x_a^2 \rangle - \langle x_a x_b \rangle \right)^2.
\end{split}
\end{equation} 

The expression in Eq.~\eqref{inversematrix} for $(\mathbb{Q}-{\bf m}
\otimes {\bf m}^T)^{-1}$ in the RS case implies that
the explicit expression of the replicon is

\begin{equation}\label{1Replic}
    \lambda_R = -(\beta J)^2 + \frac{1}{(1-q_0)^2}.
\end{equation}

For RSB saddle points, the diagonalization of
the Hessian matrix is more difficult, since the three different kinds
of elements in Eq.~\eqref{Hessian_elements} have more than just three
possible values. The computation for the fluctuations around a 1RSB
saddle point can be found in Ref.~\cite{Crisanti92}, while the
generalization to generic $k$-RSB saddle points can be found in
Ref.~\cite{Temesvari94}. But for the present case this further step of
the calculation is not necessary, since the RS ansatz turns out to be always stable.

\subsubsection{Phase diagram}

\begin{figure}  
    \includegraphics[width=\columnwidth]{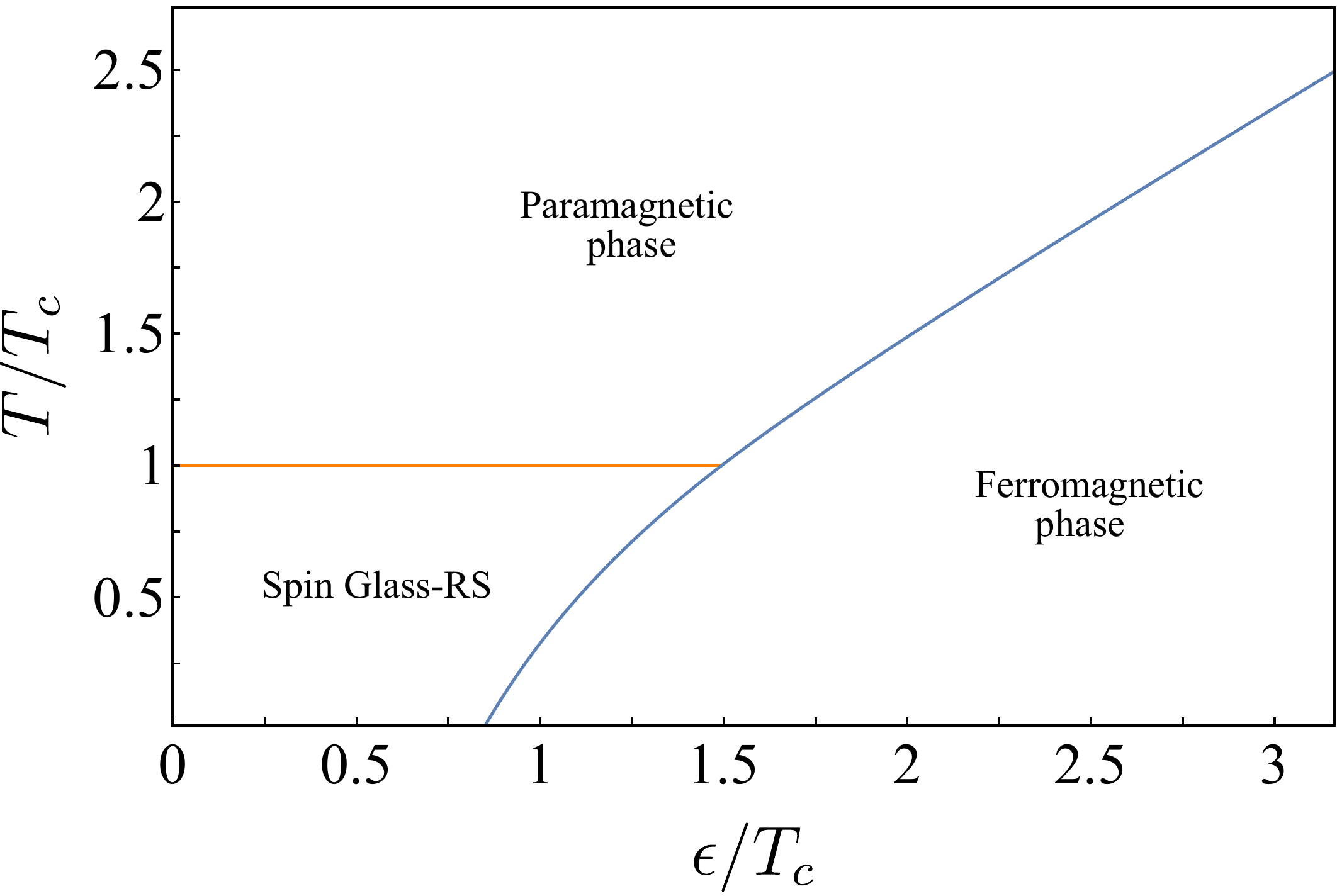}
    \caption{Phase diagram in the $(\epsilon,T)$ plane for the $p=2$
      spherical spin glass plus four-body ferromagnetic interactions,
      where $\epsilon$ represents the strength of 4-body
      ferromagnetic interactions.}
    \label{fig:PhaseDiag1}
\end{figure}

The phase diagram of the model with ordered $p=4$ interaction,
represented by the Hamiltonian in Eq.~\eqref{1_Ham}, is then obtained
by looking for the solutions of the RS saddle-point equations,
Eqns.~\eqref{1-RSspEqsALLSOL1}\eqref{1-RSspEqsALLSOL2}, and studying
their stability. The result of this analysis is presented in
Fig.~\ref{fig:PhaseDiag1}. For $T>T_c$ and small values of $\epsilon$
the system is in a paramagnetic phase, which is stable because the
replicon is always positive, as can be checked by plugging into
Eq.~\eqref{1Replic} the values $q_0=0$ and $T>T_c=J$.

Then, for small
values of $\epsilon$ by lowering the temperature below $T_c$ the
system passes to a trivial spin-glass phase, which is marginally stable, since $\lambda_R = 0$ in all this region of the parameters as can be checked by using the expression $q_0 = 1 - \frac{1}{\beta J}$ in Eq.~\eqref{1Replic}. On the contrary, if $\epsilon$ is
large enough and the non-linear ferromagnetic interaction prevails on
the disordered two-body interaction, that is for $\epsilon\gtrsim 1.5$,
upon lowering the temperature there is no spin-glass transition and
the system moves directly to the ferromagnetic phase. By using Eqs.~\eqref{1-RSspEq2} and~\eqref{1Replic} one finds the expression 
\begin{equation}
  \lambda_R = \frac{1}{4 \beta \epsilon q_0 (1-q_0)^3}
\end{equation}
from which one concludes that $\lambda_R > 0$ for all values of $q_0
\in [0,1]$. We also find a small interval of $\epsilon$ values,
approximately $\epsilon\in [0.85:1.5]$ where at $T=T_c$ the system
encounters first a transition to the (trivial) spin-glass phase and
then, upon lowering further the temperature, a transition from
(trivial) spin-glass to the ferromagnetic phase. Let us stress that in
the phase diagram of Fig.~\ref{fig:PhaseDiag1} the transition line
which separates the ferromagnetic phase from the other phases is a
line of first order transitions: the paramagnetic phase and the
trivial spin glass phase, respectively at temperatures above and below
the critical one, are always stable upon increasing $\epsilon$ and the
transition to the ferromagnetic phase is controlled by the free-energy
balance.

We can therefore conclude that the addition of ordered non-linear
interactions does not produce any sort of RSB in the trivial spin
glass phase of the spherical model. \newcol{If one considers only
  transitions of continuous kind to RSB phases, the same conclusion
  could have been drawn by considering the projection of the ordered
  perturbation on the eigenspace of the replicon. In particular, the
  eigenvector of the replicon is a vector $\delta \boldsymbol{\eta}$
  with components as in Eq.~\eqref{eq:array-fluct} where those
  corresponding to fluctuations of the magnetization are
  zero~\cite{deAlmeida78}: $\delta\boldsymbol{\eta}_R = \lbrace {\bf
    0}, \delta q_{ab}\rbrace$. Then, since an ordered perturbation can
  generate only fluctuations of the magnetization order parameter,
  $\delta m_a$, and these components live in a subspace orthogonal to
  the replicon, the ordered perturbation cannot be responsible of
  continuous transition to phases with a non-trivial RSB
  pattern. Nevertheless, since we could have in principle also
  first-order like transition to RSB phases which are not controlled
  by the stability matrix, we preferred to not only study explicitly
  the stability matrix, but to perform also a direct comparison of the
  different solutions free energy, which allowed us to draw the phase
  diagram in Fig.~\ref{fig:PhaseDiag1}.}

By comparing the present result with the phase diagram in
Fig.~\ref{fig:PhaseDiag0} we can remark a behaviour which is
topologically analogous, upon increasing the strength of the
non-linearity $\epsilon$, to that of the $p=2$ spherical model with
ferromagnetic two-body interactions competing with the disordered
ones~\cite{Kosterlitz76}: the trivial spin-glass phase remains
marginally stable until the point where a transition to ferromagnetic
order takes place.

Notice also, that this scenario holds for arbitrary ordered non-linearity, as one can see with a straightforward generalization of the previous analysis. If we denote by $p$ the power of the non-linearity, the only change in the action~\eqref{1-Action} is that the magnetization term is raised to the power $p$. As a consequence the RS saddle point equation obtained deriving the action w.r.t~$q_0$ \eqref{1-RSspEqsALLSOL1} remains the same and that obtained deriving w.r.t.~$m$ reads
\begin{gather}
   p \beta \epsilon m^{p-1} + \frac{m}{1-q_0} = 0.
\end{gather}
Therefore, the only change is in the equation for the overlap in the ferromagnetic phase, which reads
\begin{equation}
    q_0 (1-q_0)^{\frac{2(p-1)}{p-2}} \left[\frac{1}{1-q_0} - (\beta J)^2 \right] = \frac{1}{(p \beta \epsilon)^{\frac{2}{p-2}}},
\end{equation}
from which one immediately recognize that
\begin{equation}
    \lambda_R = \left(  \frac{1}{p \beta \epsilon q_0 (1-q_0)^{p-1}}  \right)^{\frac{2}{p-2}} > 0
\end{equation}
for all values of $p,\beta,\epsilon$ and $q_0\in[0,1]$.

\section{Purely disordered nonlinearity}
\label{sec:non-linear-disordered}

In this section we consider a different kind of perturbation to the
spherical $p=2$ model, namely we added a 4-body interaction term with
quenched random couplings. The model is defined by the Hamiltonian

\begin{align}
\Ha_{\text{dis}}(\boldsymbol{\sigma}) = \Ha_J(\boldsymbol{\sigma}) - \sum_{i<j<k<l}^{1,N} J^{(4)}_{ijkl}~\sigma_i\sigma_j\sigma_k\sigma_l,
\end{align}
where the couplings $J_{ijkl}$ are i.i.d. random variables following the
Gaussian distribution
\begin{align}
p(J_{ijkl}) = \frac{1}{\sqrt{2\pi\sigma_4^2}}e^{-\frac{J_{ijkl}^2}{2\sigma_4^2}},
\end{align}
with variance 
\begin{align}
\sigma_4^2 = \frac{\epsilon^2~4!}{2 N^3}.
\end{align}
By averaging over the disorder, one reaches the following form for the
replicated partition function:
\begin{align} \label{2-PartFunc}
\overline{\Z^n} &= \int \prod_{a<b}^{1,n} \de q_{ab}~e^{-N G[\mathbb{Q}]},
\end{align}
where the action $G[\mathbb{Q}]$ is defined as
\begin{equation}
\begin{aligned}
        G[\mathbb{Q}] = &- \frac{(\beta J_2)^2}{2} \sum_{a \leq b}^{1,n}
        q_{ab}^2 - \frac{(\beta \epsilon)^2}{2} \sum_{a\leq b}^{1,n} q_{ab}^4
        \\ &- \frac{1}{2} \ln\det \mathbb{Q}.
\end{aligned}
\end{equation}

The one above is the free energy of the so-called mixed $p$-spin
model, whose equilibrium thermodynamics has been studied extensively
in~\cite{Crisanti04,Crisanti06}, while more recent and very
interesting results on the dynamics can be found
in~\cite{FFT20,FFT21}. Let us notice that in the present case, at
variance with the model studied in Sec.~\ref{sec:non-linear-ordered},
due to the absence of any ferromagnetic term in the Hamiltonian there
is only one ordered parameter, the overlap matrix $\mathbb{Q}$.

\subsection{RS saddle-point equations and stability}

The RS free energy of the $2+4$-spin model, obtained
through steps analogous to those reported in App.~\ref{AppA}, reads as
\begin{equation} \label{2-RSFreeEn}
\begin{aligned}
   \lim_{n\rightarrow 0} \frac{1}{n} G[q_0] = &- \frac{(\beta J_2)^2}{4} (1-q_0^2) - \frac{(\beta \epsilon)^2}{4} (1-q_0^4)  \\
    &- \frac{1}{2}\ln(1-q_0) - \frac{1}{2} \frac{q_0}{1-q_0},
\end{aligned}
\end{equation}
from which the saddle point equation in $q_0$ is 
\begin{equation}\label{2-saddleEq}
    q_0\left[(\beta J_2)^2+2(\beta \epsilon)^2q_0^2-\frac{1}{(1-q_0^2)}\right]=0.
\end{equation}

By proceeding in a similar way with respect to Sec.~\ref{1-Stability},
we rewrite the effective action as
\begin{equation}
    \begin{aligned}
    G[\mathbb{Q}] = &- \frac{(\beta J_2)^2}{2} \sum_{a < b}^{1,n} q_{ab}^2 - \frac{(\beta \epsilon)^2}{2} \sum_{a < b}^{1,n} q_{ab}^4 \\
        & + \ln \int \prod_{a=1}^n \de x_a e^{-\sum_{a \leq b}^n q_{ab} x_a x_b },
\end{aligned}
\end{equation}
from which the expressions for the stability matrix elements read as
\begin{subequations} \label{2-PQR}
    \begin{align}
        &G_{(ab),(ab)} = -(\beta J_2)^2 - 6(\beta \epsilon)^2 q_{ab}^2 + \langle x_a^2 \rangle \langle x_b^2 \rangle + \langle x_a x_b \rangle^2 \\
        &G_{(ab),(ac)} = \langle x_a^2\rangle \langle x_b x_c \rangle + \langle x_a x_b \rangle\langle x_a x_c \rangle  \\
        &G_{(ab),(cd)}  = \langle x_a x_c \rangle \langle x_b x_d \rangle + \langle x_a x_d \rangle\langle x_b x_c \rangle,
    \end{align}
\end{subequations}
where in this case
\begin{equation}
    \langle x_a x_b \rangle = (q^{-1})_{ab}.
\end{equation}

By recalling the general expression for the replicon in the case of
RS ansatz, given in Eq.~\eqref{1-replicon}, we obtain
in the present case

\begin{equation}\label{2-RSrepl}
    \lambda_R = -(\beta J_2)^2 - 6(\beta \epsilon)^2 q_0^2 + \frac{1}{(1-q_0)^2}.
\end{equation}

By plugging the expression of Eq.~\eqref{2-saddleEq} into
Eq.~\eqref{2-RSrepl} it is easy to realize that $\lambda_R$ is
negative for any value of the parameter $q_0$ different from zero, so
that in the RS case only the solution $q_0=0$ can be
stable. By then plugging the latter into the expression of the
replicon we get $\lambda_R=1-(\beta J_2)^2$, which for temperatures
$T<T_c=J_2$ becomes negative, signalling the instability of the RS
ansatz and therefore forcing us to consider RSB
solutions.

\subsection{1RSB saddle-point equations}\label{1rsb-2}

The first attempt to go beyond a RS ansatz it
is always represented by considering one step of replica symmetry
breaking (1RSB), corresponding to an overlap matrix, which we may refer
to as $\mathbb{Q}^{\text{1step}}$, with the following structure:
\begin{align}
[\mathbb{Q}^{\text{1step}}]_{ab} = (1-q_1)\delta_{ab} +(q_1-q_0)\epsilon_{ab}+ q_0 \mathbb{I}_{ab}.
\end{align} 
$\mathbb{I}={\bf 1}\otimes {\bf 1}^T$ is a matrix whose elements are all identically equal to $1$, while $\epsilon$ is a block diagonal matrix, with diagonal blocks all equal to $\mathbb{I}_x={\bf
  1}_x\otimes {\bf 1}_x^T$, namely $x\times x$ square matrices with
all elements identically equal to one. For the 1RSB ansatz we have therefore three variational
parameters: $q_0$, $q_1$ and $x$. In the present case the inverse
matrix of matrix $q_{ab}$ reads as:

\begin{align}
[(\mathbb{Q}^{\text{1step}})^{-1}]_{ab} = A\delta_{ab} +B\epsilon_{ab}+ C \mathbb{I}_{ab},
\end{align}
with
\begin{gather}
    A=\frac{1}{1-q_1}  \\
    B=-\frac{(q_1-q_0)}{(1-q_1)(1-q_1+x(q_1-q_0))} \\
    C=-\frac{q_0}{[1-q_1+x(q_1-q_0)]^2}.
\end{gather}
Given this structure of the overlap matrix, with very similar
calculations to the RS case and defining the function
\begin{equation}\label{chi-notation}
\chi_p=\chi_p(q_0,q_1,x)\equiv 1-q_1^p+x(q_1^p-q_0^p),
\end{equation}
we get as 1RSB effective action:
\begin{align}
  &\lim_{n\rightarrow 0} \frac{1}{n} G[q_0,q_1,x] = -\frac{(\beta
    J_2)^2}{4}\chi_2 -\frac{(\beta \epsilon)^2}{4}\chi_4- \nonumber
  \\ & \frac{x-1}{2x}\log(1-q_1)-\frac{1}{2x}\log(\chi_1)-\frac{1}{2}\frac{q_0}{\chi_1}.\notag
\end{align}

\begin{comment}
    \begin{align}
    \lim_{n\rightarrow 0} \frac{1}{n} G[q_0,q_1,x] = &-\frac{(\beta J_2)^2}{4}[1-q_1^2 + x(q_1^2-q_0^2)]+\\
    & -\frac{(\beta \epsilon)^2}{4}[1-q_1^4 + x(q_1^4-q_0^4)] -\frac{x-1}{2x}\log(1-q_1) + \notag\\
    & -\frac{1}{2x}\log(1-q_1 + x(q_1-q_0)) -\frac{1}{2}\frac{q_0}{1-q_1+x(q_1-q_0)}.\notag
\end{align}
\end{comment}

The 1RSB saddle point equations read as:
\begin{widetext}
   \begin{subequations} \label{1RSBsaddleeqs2}
    \begin{align}
        &\frac{\partial G}{\partial q_0} = \frac{x}{2}q_0\left[(\beta J_2)^2+2(\beta )^2 q_0^2-\frac{1}{\chi_1^2}\right]=0,\\
        &\frac{\partial G}{\partial q_1} = (x-1)\left[-q_1 (\beta J_2)^2-2 q_1^3 (\beta \epsilon)^2+\frac{q_0}{\chi_1^2} +\frac{q_1-q_0}{(1-q_1)\chi_1}\right]=0,\\
        &\frac{\partial G}{\partial x} = -\frac{(\beta J_2)^2}{2}[q_1^2-q_0^2]-\frac{(\beta \epsilon)^2}{2}[q_1^4-q_0^4]+\frac{1}{x^2} \log\frac{\chi_1}{1-q_1} +(q_1-q_0)\left[\frac{q_0}{\chi_1^2}-\frac{1}{x\cdot\chi_1}\right]=0.
    \end{align}
\end{subequations} 
\end{widetext}

\subsection{Stability of the 1RSB saddle point and phase diagram} \label{2-Stability}
For the diagonalization of the stability matrix of the 1RSB ansatz we
have followed the procedure of~\cite{Crisanti92}, which generalizes
RS calculations of~\cite{deAlmeida78}.

\subsubsection{Replicon}

There are three different classes of fluctuations around the 1RSB
saddle point, leading to nine different eigenvalues, which reduce to
seven in the limit $n \rightarrow
0$. Following~\cite{Crisanti04,Crisanti06}, the two relevant
eigenvalues for the study of the stability are

\begin{gather}
    \Lambda^{(1)} = -(\beta J_2)^2 - 6(\beta \epsilon)^2 q_1^2 + \frac{1}{(1-q_1)^2}\label{lamb1}  \\
    \Lambda^{(2)} = -(\beta J_2)^2 - 6(\beta \epsilon)^2 q_0^2 + \frac{1}{(1- q_1+x (q_1-q_0))^2}\label{lamb2}.
\end{gather}

Fluctuations with respect to a given ansatz for the breaking of
replica symmetry means fluctuations which alter the structure of the
matrix $\mathbb{Q}^{\text{1step}}$. In general for these instabilities
it is also possible to give a physical interpretation connected to
the modification of the matrix structure. In particular, in the case
of a block diagonal matrix $\mathbb{Q}^{\text{1step}}$, each row of
the matrix has the same structure: one element is unity, there are
$x-1$ elements equal to $q_1$, those belonging to the same block
$\mathbb{I}_x$, and $n-x$ elements equal to $q_0$. This structure of
the matrix corresponds to a {\it ``one-step''} fragmentation of phase
space in disjoint ergodic components, which we may refer to as {\it
  states} or {\it clusters}. Configurations belonging to the same
state have typical overlap $q_1$, while configurations
belonging to different states have overlap $q_0$. By
considering the structure of the matrix $\mathbb{Q}^{\text{1step}}$, a
consistent interpretation is to regard the elements of blocks
$\mathbb{I}_x$ as the typical overlap between configurations belonging
to the same state and the elements outside blocks
$\mathbb{I}_x$ as the typical overlap between configurations belonging
to different states. 

Given this scenario, we consider two kind of instabilities: 1) a single state can
undergo a further fragmentation process, which corresponds to an emerging
block structure {\it inside} $\mathbb{I}_x$, with new elements $q_2$
appearing within $\mathbb{I}_x$ (again in a block-diagonal structure);
2) different states can merge into a single one, which corresponds to a
rearrangement of the structure of the whole
$\mathbb{Q}^{\text{1step}}$, with {\it all} inner elements of
different blocks $\mathbb{I}_x$ changing value from $q_1$ to $q_2$ and
{\it some} of the extra-block elements rising from $q_0$ to $q_1$.
These two patterns to alter the 1RSB structure of
$\mathbb{Q}^{\text{1step}}$ are connected respectively to the
eigenvalues $\Lambda^{(1)}$ and $\Lambda^{(2)}$. In particular, the fluctuations inside the same cluster
represented by $\Lambda^{(1)}$ are the ones which destabilize the 1RSB
phase in the SK model and that lead, across an
infinite sequences of further other breakings, to the well known
fractal free-energy structure~\cite{Parisi79}. This instability
pattern for the 1RSB phase is the most common: it is for instance the
one which is also found in models of hard
spheres~\cite{CKPUZ14b,PaUrZa20} and ecological
models~\cite{Altieri21}. 

On the other hand, for mixed $p$-spin models, i.e.,
models with spherical variables and competition between linear and
non-linear interactions, it turns out the dominant instabilities are
those related to $\Lambda^{(2)}$, as demonstrated in~\cite{Crisanti06}
by showing that $\Lambda^{(1)}>\Lambda^{(2)}$, so that the latter is
the dominant eigenvalue. This means that the \textit{relevant}
replicon is the one related to fluctuations between clusters. In this
situation it was pointed out in~\cite{Crisanti06} that an intermediate
phase also emerges, the so-called {\it 1-full-RSB} phase,
characterized by the coexistence between a one-step and full replica
symmetry breaking (full-RSB). In the present discussion we will not deem to
study in detail the transition between this mixed 1-full-RSB phase and the pure full-RSB phase, which also takes place in the
present case, referring the interested reader to~\cite{Crisanti06}. 

\subsubsection{Phase diagram}

In the mixed $p$-spin with purely disordered nonlinearity we keep
fixed the variance of the two-body disordered interaction, as in the
previous study of Sec.~\ref{sec:non-linear-ordered}, and study the
phase diagram in the plane $(\epsilon,T)$, where in the present case
$\epsilon$ represents the standard deviation of the four-body
couplings disorder. As in the previous case, we find that the lower
limit of stability of the paramagnetic phase is $T_c = J_2$. For
$\epsilon$ high enough one finds at all temperatures a transition to a
1RSB glassy phase. Then, depending on the temperature, upon lowering $\epsilon$
we can either find a transition to the paramagnetic phase, for $T>T_c$,
or a transition to a full-RSB phase, for
$T<T_c$, (either {\it 1-full-RSB} or standard full-RSB). 

Let us remark the difference between the transition taking place from
the 1-RSB phase to the paramagnetic one above $T_c$ and the one taking
place from the 1-RSB to the full-RSB below $T_c$. Considering the
probability distribution of the overlap between replicas, which
corresponds to the parametrization of the matrix
$\mathbb{Q}^{\text{1step}}$, the transition at temperatures $T>T_c$
has the features of a first-order phase transition, and is in fact
usually known as ``Random First-Order Transition''~\cite{C08,BB11},
while the transition at temperatures $T<T_c$ has the features of a
second order phase transition. This means that for $T>T_c$, upon
reducing $\epsilon$, the 1RSB ansatz for $\mathbb{Q}^{\text{1step}}$
is always a {\it locally stable} solution, even when it becomes less
convenient than the RS ansatz from the point of view of free energy,
until when it completely disappears as a solution at the line marked
in the phase diagram as {\it 1RSB spinodal}. This is the transition
mechanism typical of first-order transitions. Technically, the
spinodal line is simply obtained looking for the value of $\epsilon$
where non-trivial solutions of the equation $\partial G/\partial
q_1=0$ arise in terms of $q_1$ with $x=1$ and $q_0=0$ fixed, for
instance by increasing $\epsilon$ at fixed temperature $T$.

\begin{figure} 
    \includegraphics[width=\columnwidth]{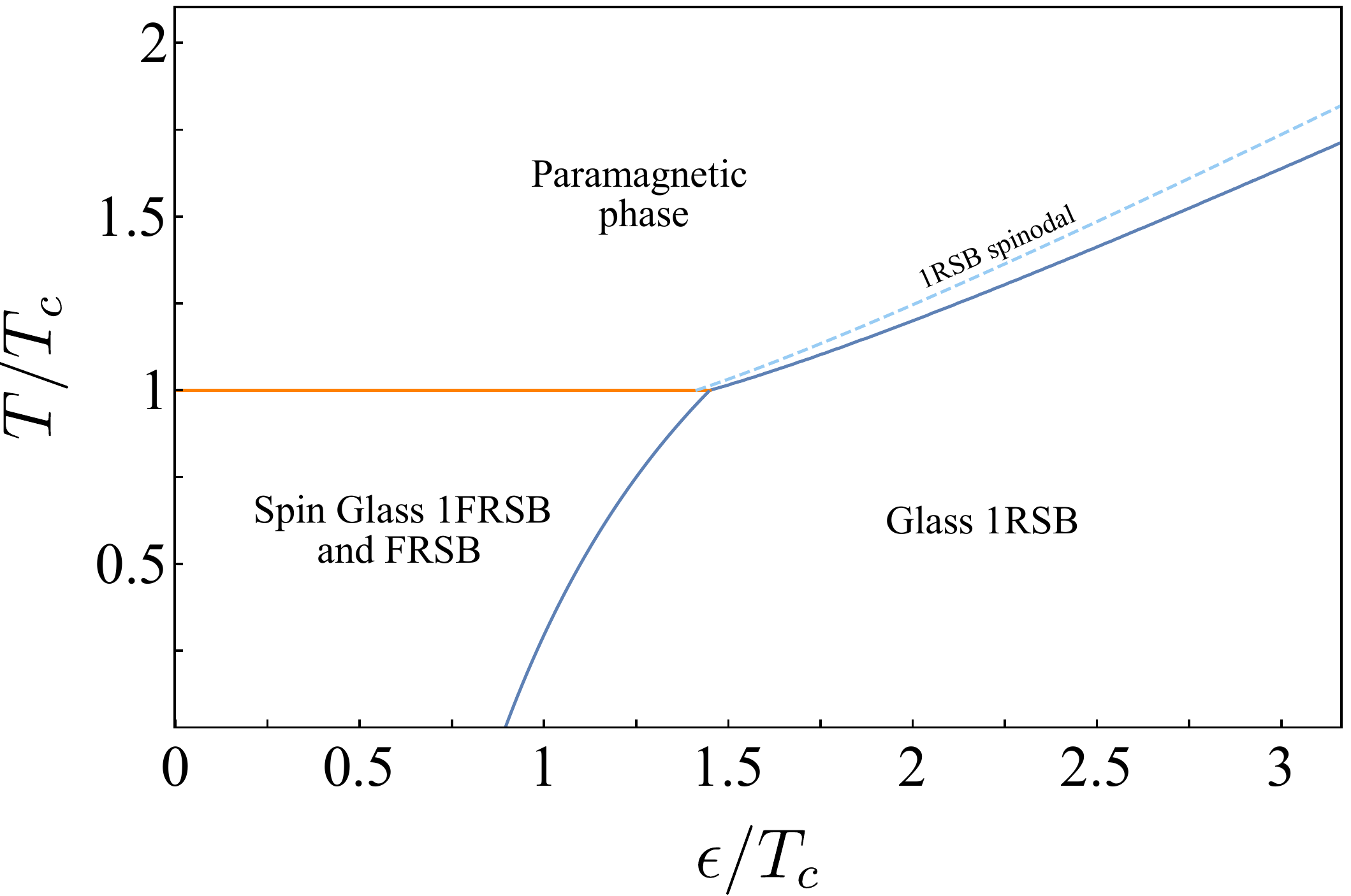}
    \caption{Phase diagram in the $(\epsilon,T)$ plane for the $p=2$
      spherical spin glass plus four-body {\it purely disordered}
      interactions, where $\epsilon$ represents the strength of
      non-linear interactions.}
    \label{fig:PhaseDiag2}
\end{figure}

On the other hand, for $T<T_c$, the transition from the 1RSB phase to
the full-RSB phase takes place when the former loses stability in
favour of the latter, as happens typically in second order phase
transitions. The transition line in Fig.~\ref{fig:PhaseDiag2} has been
obtained by finding the values of the parameters where $\Lambda^{(2)}
= 0$. Despite this difference concerning the behaviour of the overlap
matrix, both transitions are continuous from the point of view of
Landau classification, because in both cases there is no latent heat.

\section{Ferromagnetic+disordered non-linearity}
\label{sec:non-linear-mixed}

Having assessed the effect of adding respectively purely ordered and
purely disordered non-linear terms to the $p=2$ spherical model, in this
section we study the effect of non-linear terms with both competing
disordered and ferromagnetic interactions. This can be done by
extracting the couplings of the 4-body interaction term from a
non-zero average probability distribution, namely we consider

\begin{equation}
    p(J_{ijkl}) = \frac{1}{\sqrt{2\pi\sigma_4^2}}e^{-\frac{(J_{ijkl}-J_0)^2}{2\sigma_4^2}},
\end{equation}
with
\begin{equation}
    \sigma_4^2 = \frac{\epsilon^2~4!}{2 N^{3}}~~~~~~~J_0 = \frac{\epsilon^2~4!}{N^3}.
    \label{eq:eps-dis-ferro}
\end{equation}

The variable $\epsilon$ parametrizes both the mean and the variance of
the 4-body couplings distribution. This choice allows to have just one
free parameter for the whole Gaussian distribution. With a similar computation with respect to the previous cases we find the effective
action
\begin{equation}
\begin{split}
        G[\mathbb{Q},{\bf m}] = &- \frac{(\beta J)^2}{2} \sum_{a<b}^{1,n} q_{ab}^2 - \frac{(\beta \epsilon)^2}{2} \sum_{a < b }^{1,n} q_{ab}^4 \\
        & - \beta\epsilon^2  \sum_{a=1}^n m_a^4 - \frac{1}{2} \ln\det\left(\mathbb{Q}-{\bf m} \otimes {\bf m}^T\right).
\end{split}
\end{equation}

\subsection{RS solutions}

In the same way of previous sections we begin by considering a
RS ansatz for the matrix $\mathbb{Q}$. This leads to a
RS action reading as:

\begin{align}
    \lim_{n\rightarrow 0} \frac{1}{n} G[q_0,&m]=-\frac{(\beta J_2)^2}{4}(1-q_0^2)-\frac{(\beta\epsilon)^2}{4}(1-q_0^4) \nonumber \\
    &-\beta\epsilon^2 m^4 -\frac{1}{2}\log(1-q_0)-\frac{1}{2}\frac{q_0-m^2}{1-q_0}.
\end{align}

The RS saddle-point equations read as:

\begin{align}
  \frac{\partial G}{\partial q_0}& = (\beta J_2)^2 q_0+2 \beta^2 \epsilon^2 q_0^3-\frac{q_0-m^2}{(1-q_0)^2}=0 \nonumber \\
  \frac{\partial G}{\partial m}& = -4\beta \epsilon^2 m^3+\frac{m}{1-q_0}=0
  \label{3-RSsaddleeqs}
\end{align}

There are three possible solutions for Eq.~\eqref{3-RSsaddleeqs}:

\begin{enumerate}
    \item $m=0$;\\ $q_0=0$.
    \item $m=0$;\\ $(\beta J_2)^2+2\beta^2 \epsilon^2  q_0^2 = 1/(1-q_0)^2$.
    \item $m=1/\sqrt{4 \beta \epsilon^2(1-q_0)}$;\\
      $q_0 (1-q_0)[1-(1-q_0)^2((\beta J_2)^2+2 \beta^2 \epsilon^2
        q_0^2)]=1/(4 \beta \epsilon^2)$.
\end{enumerate}

\begin{figure} 
    \centering
    \includegraphics[width=\columnwidth]{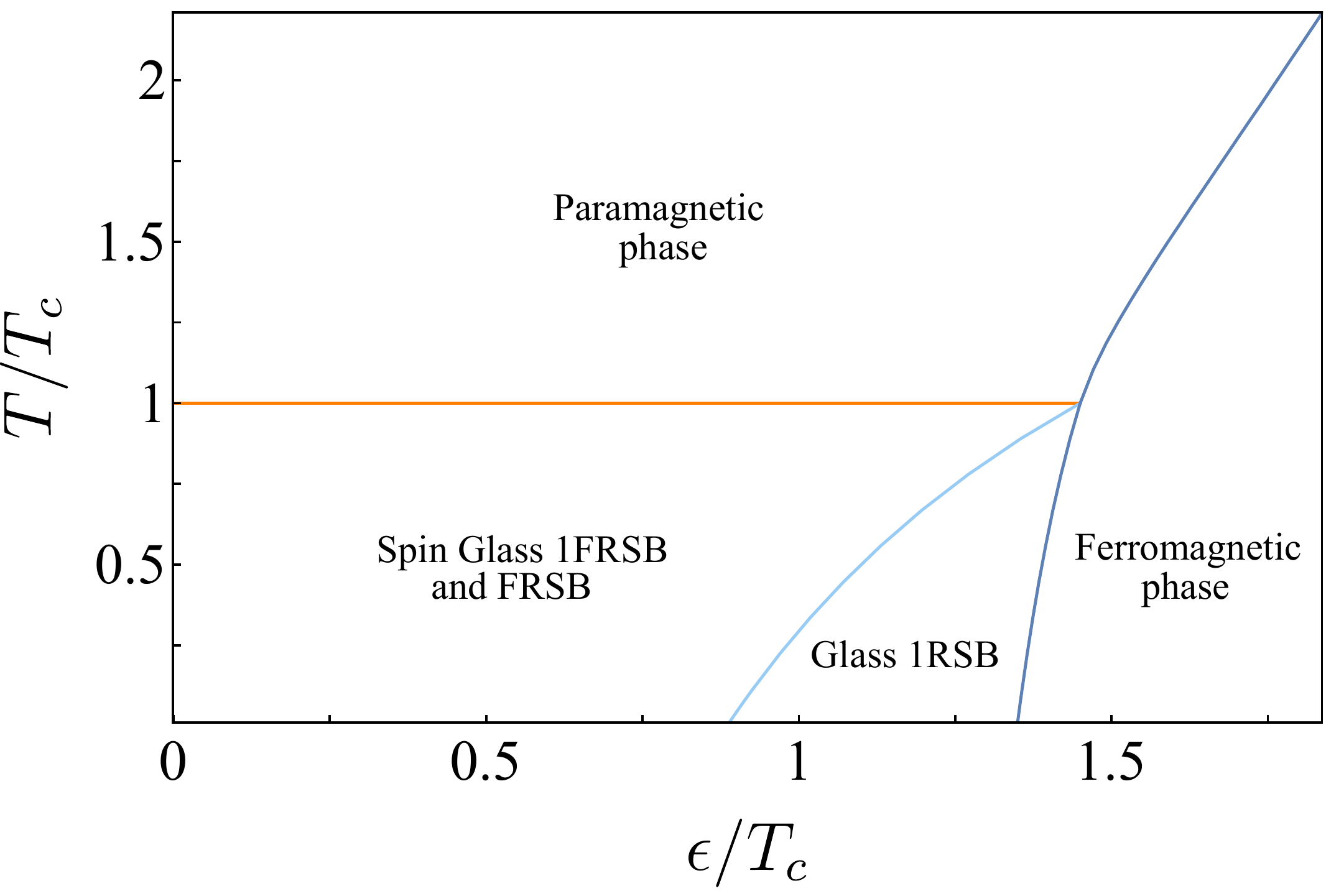}
    \caption{Phase diagram in the $(\epsilon,T)$ plane for the $p=2$
      spherical spin glass plus four-body interactions, with competing
      disordered and ferromagnetic terms, where the variance of the
      disorder and the strength of ferromagnetic couplings are both
      controlled by $\epsilon$, see Eq.~\eqref{eq:eps-dis-ferro} in the text.}
    \label{fig:PhaseDiag3}
\end{figure}

In order to study the stability of the solution we have computed the
replicon, which for the present case reads as:
\begin{equation}
    \lambda_R=\frac{1}{(1-q_0)^2}-(\beta J_2)^2
-6 \beta^2 \epsilon^2 q_0^2.
\end{equation}
The paramagnetic solution (PM, n.1) with $m=0$ and $q_0=0$ is unstable everywhere
below the critical line $T_c=J_2$, since $\lambda_R>0$. Therefore, below this line, which is
the horizontal orange one in the phase diagram of
Fig.~\ref{fig:PhaseDiag3}, we can only have phases with either finite
magnetization or replica symmetry breaking or both. The trivial spin glass solution (n.2), is never stable for this model because one always has $\lambda_R<0$,
therefore it is always excluded from the phase diagram. For what concerns
the ferromagnetic phase (FM, n.3) in the list above, the numerical solution of the
corresponding saddle-point equations shows that in the range of
parameters $(\epsilon,T)$ where such a solution exists and is
nontrivial, it is stable, i.e.~$\lambda_R > 0$. 

Regarding the PM-FM transition, since both solutions are always stable, it is only by
comparing their free energies that we can determine the transition line. The
coordinates of this first-order transition line in the phase diagram in Fig.~\ref{fig:PhaseDiag3} are obtained by the implicit equation: 
\begin{align}
  f^{\text{para}}(\epsilon_{c}^{\text{ferro}},T_{c}^{\text{ferro}}) = f^\text{ferro}(\epsilon_{c}^{\text{ferro}},T_{c}^{\text{ferro}}),
\end{align}
where:
\begin{align}
  \text{sol n.1} ~~\Longrightarrow ~~ & f^{\text{para}}(\epsilon,T) = \frac{1}{\beta} G(q_0=0,m=0) \nonumber \\
  & = -\frac{\beta}{4}(J_2^2+\epsilon^2)  \nonumber \\
\text{sol n.3} ~~\Longrightarrow ~~ & f^{\text{ferro}} = G(q_0>0,m>0).
\end{align}

We will comment the remaining part of the phase diagram, namely what
happens at temperatures lower than the critical one for the stability
of the paramagnetic phase, $T<T_c$, in the following paragraph after
having introduced the 1RSB ansatz for the replica matrix.

\subsection{1RSB ansatz}

In order to study the phase diagram of the model for $T<T_c$ we need to consider the possibility of replica symmetry breaking. By recalling the definition

\begin{align}
  \chi_p(q_0,q_1,x) &= 1-q_1^p+x(q_1^p-q_0^p)
\end{align}
and by assuming the 1RSB structure for the matrix $\mathbb{Q}$ we obtain
the following form for the effective action in the limit $n\rightarrow 0$:

\begin{widetext}
  
\begin{align}
  \lim_{n\rightarrow 0} \frac{1}{n} G[q_0,q_1,x,m]= -\newcol{\frac{(\beta J_2)^2}{4}}\chi_2(q_0,q_1,x)-\frac{\beta^2 \epsilon^2}{4}\chi_4(q_0,q_1,x) -\beta \epsilon^2 m^4-\frac{x-1}{2x}\ln(1-q_1)-\frac{\ln(\chi_1)}{2x}-\frac{q_0-m^2}{2 \chi_1(q_0,q_1,x)}, \nonumber \\
\end{align}
from which the four 1RSB saddle point equations read as:
\begin{align}
  \frac{\partial G}{\partial q_0} &=\frac{(\beta J_2)^2}{2}x q_0+ \beta^2 \epsilon^2 x q_0^3-\frac{1}{2}\frac{(q_0-m^2) x}{\chi_1^2(q_0,q_1,x)}=0 \nonumber \\
  \frac{\partial G}{\partial q_1} &= (x-1)\bigg[-\frac{(\beta J_2)^2}{2}q_1-\beta^2 \epsilon^2 q_1^3+\frac{1}{2 x(1-q_1)} -\frac{1}{2 x\chi_1(q_0,q_1,x)}+\frac{q_0-m^2}{2\chi_1^2(q_0,q_1,x)}\bigg]=0 \nonumber \\
  \frac{\partial G}{\partial x} &= -\frac{(\beta J_2)^2}{4}(q_1^2-q_0^2)-\frac{\beta^2 \epsilon^2}{4}(q_1^4-q_0^4)-\frac{1}{2}\frac{q_1-q_0}{x \chi_1(q_0,q_1,x)} +
  \frac{1}{2 x^2}\ln\left(1+\frac{x(q_1-q_0)}{1-q_1}\right)+\frac{(q_0-m^2)(q_1-q_0)}{2 \chi_1^2(q_0,q_1,x)}=0 \nonumber \\
  \frac{\partial G}{\partial m} &= m \left(-4\beta \epsilon^2 m^2+\frac{1}{\chi_1(q_0,q_1,x)}\right)=0.
  \label{1RSBsaddleeqs}
\end{align}

The expression of the two replicon eigenvalues which are relevant in the limit $n\rightarrow 0$,
which can be obtained following the lines of~\cite{Crisanti06}, reads as
\begin{gather}
    \Lambda^{(1)}[q_1] = -(\beta J_2)^2 - 6\beta^2 \epsilon^2 q_1^2 + \frac{1}{(1-q_1)^2},  \\
    \Lambda^{(2)}[q_0,q_1,x]= -(\beta J_2)^2 - 6\beta^2 \epsilon^2 q_0^2 + \frac{1}{(1- q_1+x (q_1-q_0))^2}.
\end{gather}
\end{widetext}

As already mentioned, for this model the most relevant between the two
is $\Lambda^{(2)}$, signalling a pattern of instability of the 1RSB ansatz characterized
by the merging of 1RSB clusters.

\subsubsection{Phase Diagram}
Let us here summarize how all the transition lines for $T<T_c$ in Fig.~\ref{fig:PhaseDiag3} have been obtained. We find, for $\epsilon$ large enough, a stable 1RSB phase
with zero magnetization, $m=0$. The transition line which separates
this 1RSB phase, which we refer to simply as {\it glass} phase, from
the spin-glass phase (with full-RSB) at smaller value of $\epsilon$ has been determined by simply tracking the value of the control parameters where the 1RSB phase loses stability. That is, the intrinsic
equation of this line is determined as
\begin{equation}
\Lambda^{(2)}[q_0(\epsilon_c^{sg},T_c^{sg}),q_1(\epsilon_c^{sg},T_c^{sg}),x(\epsilon_c^{sg},T_c^{sg})] = 0.
\end{equation}
Inside the region of the phase diagram delimited by the lines $T=T_c$
above and $(\epsilon_c^{sg},T_c^{sg})$ to the right, both the RS and
the 1RSB solutions are unstable. This is therefore necessarily a
region where further breakings of the symmetry between replica
occurs. As already mentioned, we know from the results
of~\cite{Crisanti04} that different kinds of full-RSB occur here. Both
the transitions occourring along these lines are continuous.

Finally, the transition line at $T<T_c$ between the RS ferromagnetic
phase at large $\epsilon$ and the intermediate zero magnetization 1RSB
phase at smaller values of $\epsilon$ is determined by simply
comparing their free energies and has the implicit equation:
\begin{align}
f^{\text{ferro}}(\epsilon_c^{\text{ferro/1rsb}},T_c^{\text{ferro/1rsb}}) = f^{\text{1rsb}}(\epsilon_c^{\text{ferro/1rsb}},T_c^{\text{ferro/1rsb}}).
\end{align}
This is clearly a first order transition.

\begin{figure}  
  \includegraphics[width=\columnwidth]{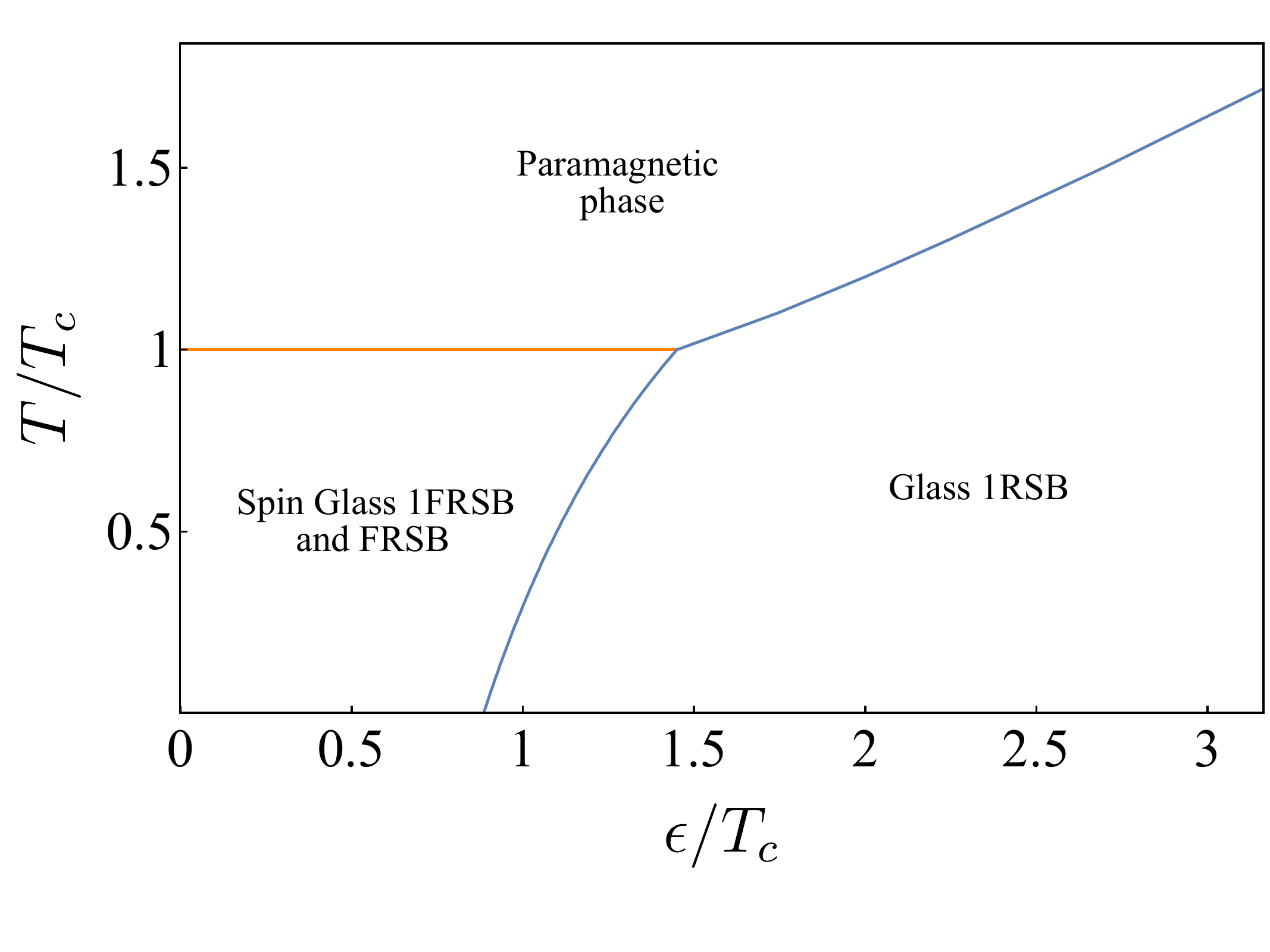}
  \caption{\newcol{Phase diagram in the $(\epsilon,T)$ plane for the $p=2$
    spherical spin glass plus four-body interactions: Gaussian
    distribution of non-linear couplings with standard deviation
    $\sigma_4 \sim \epsilon$ and mean $J_0 \sim \epsilon$
    ($R=\text{const}$).}}
    \label{fig:Rconst}
\end{figure}

\begin{figure}  
    \includegraphics[width=\columnwidth]{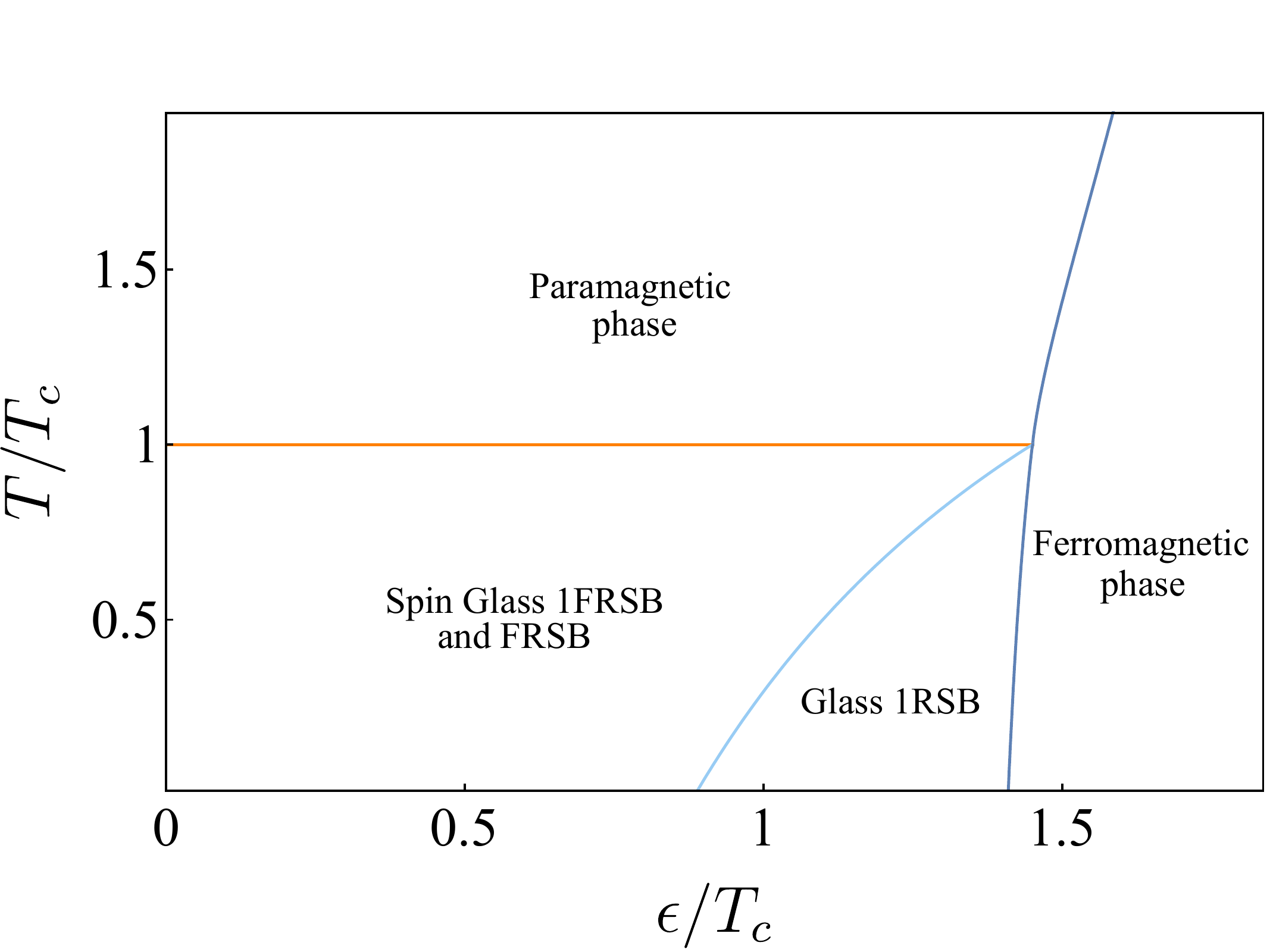}
    \caption{\newcol{Phase diagram in the $(\epsilon,T)$ plane for the $p=2$
      spherical spin glass plus four-body interactions: Gaussian
      distribution of non-linear couplings with standard deviation
      $\sigma_4 \sim \epsilon$ and mean $J_0 \sim \epsilon^4$ ($R\sim
      1/\epsilon^3$).}}
    \label{fig:Rzero}
\end{figure}

\subsubsection{\newcol{Other choices of the ratio $\sigma_4/J_0$}}

\newcol{It is legitimate to wonder whether the choice
\begin{align}
  \sigma_4/J_0=R(\epsilon) \sim \epsilon^{-1}
\end{align}
considered so far for the Gaussian distribution of non-linear
couplings is fully representative of the case where for the non-linear
couplings we have a competition between frustrated and ferromagnetic
interactions. In general for the ratio $R(\epsilon)$ we have three
possibilities: $R(\epsilon)$ is constant with respect to the parameter
$\epsilon$, which means that ferromagnetic and frustrated interactions
are always competing; $R(\epsilon)\rightarrow\infty$ when
$\epsilon\rightarrow\infty$, which means that in the large-$\epsilon$
limit frustration is dominating; $R(\epsilon)\rightarrow 0$ when
$\epsilon\rightarrow 0$, which means that in the large-$\epsilon$
limit the ferromagnetic ordering effect is dominating.}

\newcol{In order to
have a clearer picture on how the phenomenology of the model depends on
the choice of $R(\epsilon)$ we have studied the phase diagram,
repeating the analysis already discussed above, also in the case where
$\sigma_4\propto \epsilon$ and $J_0\propto \epsilon^4$, i.e., $R
\propto \varepsilon^{-3}$, to check whether the increase of
ferromagnetic terms relative importance introduces new glass phases
(for instance the ferro-1RSB phase found in~\cite{CL2013}), and in the
case $R(\epsilon) =\text{const}$, in particular assuming $\sigma_4
\propto \epsilon$ and $J_0 \propto \epsilon$.}

\newcol{We have obtained phase diagrams qualitatively identical to
  those discussed in the previous section, but it is nevertheless
  worth to report them to clarify the role played by the choice of
  $R(\epsilon)$. Let us start from the case $R(\epsilon)
  =\text{const}$, for which the phase diagram is shown in
  Fig.~\ref{fig:Rconst}: it can be easily seen that it has the same
  topology of the phase diagram of the purely frustrated case in
  Fig.~\ref{fig:PhaseDiag2}, which can be formally identified with the
  case $R=\infty$, since there we have $J_0=0$. In both cases there is
  no ferromagnetic phase! We have therefore checked, even without a
  full detailed analysis of the three-dimensional $(T,\sigma_4,J_0)$
  phase diagram, that already a constant ratio between $\sigma_4$ and
  $J_0$ is enough to produce the phenomenology that we find also in
  the limit $R=\infty$: the complete disappearance of the ordered
  ferromagnetic phase. It can be easily argued that the same will
  happen for every choice of $R(\epsilon)$ which is an increasing
  function of $\epsilon$. On the other hand we have found that the
  phase diagram of the case $R \sim \epsilon^{-3}$, which is shown in
  Fig.~\ref{fig:PhaseDiag1}, is qualitatively identical to the case $R
  \sim \epsilon^{-1}$ discussed shortly above, i.e., we do find
  ferromagnetic ordering at large $\epsilon$ and the same glassy
  phases at smaller values of the order parameter. We can therefore
  conclude that, despite the choice $R\sim \epsilon^{-1}$ of the
  previous section looks quite arbitrary, is indeed representative of
  the model phenomenology when the importance of the ferromagnetic
  terms grows with $\epsilon$.  }

\section{Discussion and conclusions}
\label{sec:conclusions}

Motivated by the idea of investigating the competition between
disorder and non-linearity, we have analytically studied the effects
of adding different kinds of non-linear interactions to the $p=2$
spherical model. In particular we studied which of these perturbations
is actually able to induce a non-trivial RSB
pattern in the low-temperature trivial spin-glass phase of the model,
which is known to be marginally stable.  We have assessed that ordered
non-linear interactions with ferromagnetic couplings are not effective
in inducing any sort of replica symmetry breaking at $T<T_c$ in the
spherical model. With respect to this kind of disturbance the
marginally stable trivial spin glass phase is in practice {\it
  stable}. The phase diagram in the plane $(\epsilon,T)$, where
$\epsilon$ is the strength of the quartic ordered non-linearity,
although the shape of the lines is slightly different, has precisely
the same topology and the same phases of the phase diagram for the
spherical spin glass with ferromagnetic couplings competing with the
disordered ones: at small values of temperature and strength of
ferromagnetic couplings we always have a trivial spin-glass phase.

By then inserting disordered four-body couplings, either purely
disordered ones as in~\cite{Crisanti06} or disordered ones competing
with ferromagnetic ones, we always find that the trivial spin-glass
phase of the spherical model is really {\it marginally} stable: any
finite strength of the additional non-linearity is able to induce a
non-trivial RSB pattern, of a kind which depends on the strength of
the disorder or on the relative importance of purely disordered and
ferromagnetic couplings. The most interesting finding concerns the
model with non-linear interactions
$J_{ijkl}\sigma_i\sigma_j\sigma_k\sigma_l$ and non-zero mean Gaussian
distribution for $J_{ijkl}$, for which we devised the following
evolution pattern below the $T_c$. By increasing simultaneously the
variance and the mean of the disorder distribution, both controlled by
$\epsilon$, we find that the model immediately sets in a full-RSB
phase, hence the trivial spin-glass phase at $\epsilon=0$ is unstable
towards a phase characterized by a hierarchical free-energy
landscape. Then, when $\epsilon$ is increased enough, the pattern of
replica symmetry breaking simplifies to a one-step hierarchy, but
still with zero magnetization. Finally, by increasing further
$\epsilon$, we have that the physics is completely dominated the
ferromagnetic interactions and there is a first-order transition to an
ordered ferromagnetic phase.

One among the pedagogical aspects of this presentation has been the
emphasis on the stability of the 1RSB phase for both the cases with
non-linear disordered couplings. In particular, it is worth mentioning
the fact that, as outlined first in~\cite{Crisanti06}, at variance
with many important models, e.g., SK model, hard spheres in infinite
dimensions and ecological models, in the $2+4$ spherical model the
instability of the 1RSB phase follows a different pattern, which is
characterized by the merging of 1RSB states in metabasins rather than
a tendency to fragmentation in smaller ones.\\

\section{Acknowledgments}
We thank M. C. Angelini, S. Franz, L. Leuzzi, T. Rizzo, and
A. Vulpiani for useful discussions. G.G. and J.N. thanks the Physics
Department of ``Sapienza'' University for kind hospitality during some
stages in the preparation of this work. G.G. and T.T.  acknowledge
partial support from the project MIUR-PRIN2022, {\it ``Emergent
  Dynamical Patterns of Disordered Systems with Applications to
  Natural Communities''}, code 2022WPHMXK.

\newpage
\appendix
\section{Details of the computation for the case of ordered non-linearity} \label{AppA}
In this Appendix we present in detail the computation of the quenched free energy of the spherical 2-spin model with ordered 4-body non-linearity.
\subsection{Computation of the quenched free energy} 
The computation of the quenched free energy with the replica method requires the knowledge of the replicated partition function, which is given by
\begin{align} \label{A1-PartFunc1}
\begin{aligned}
    \Z_J^n =  \int \D\sigma  &\exp\left[\beta \sum_{i<j}^{1,N} \sum_{a=1}^n J_{ij}\sigma_i^a\sigma_j^a\right] \\
      &\times \exp\left[\frac{4!\beta\epsilon}{N^3}\sum_{i<j<k<l}^{1,N}\sum_{a=1}^n\sigma_i^a\sigma_j^a\sigma_k^a\sigma_l^a \right],
\end{aligned}
\end{align}
where we have introduced the shorthand notation
\begin{equation} \label{A1-Desigma}
   \D\sigma = \prod_{i=1}^N\prod_{a=1}^n \de \sigma_i^a \delta\left(N-\sum_{i=1}^N (\sigma_i^a)^2\right)
\end{equation}
for the integration element over the replicated hypersphere.

By considering the $J$-dependent part of the expression \eqref{A1-PartFunc1}, we perform the average over the coupling distribution as follows 
\begin{align*}
    \overline{e^{\beta \sum_{i<j}^N \sum_a^n J_{ij}\sigma_i^a\sigma_j^a}} &= \prod_{i<j}^{1,N} \int \frac{\de J_{ij}}{\sqrt{2\pi\sigma_J^2}}e^{-\frac{J_{ij}^2}{2\sigma_J^2}} e^{\beta \sum_a^n \sigma_i^a\sigma_j^a J_{ij}} \\
    &= \exp\left[\frac{(\beta J)^2}{2N} \sum_{i<j}^{1,N} \sum_{ab}^{1,n} \sigma_i^a\sigma_i^b \sigma_j^a\sigma_j^b \right].
\end{align*}
We now retain only the leading contributions to the free energy density in the large $N$ limit, i.e.~only terms of order $O(1)$. In the expression
\begin{align*}
\begin{aligned}
    2 \sum_{i<j}^{1,N} \sum_{ab}^{1,n} \sigma_i^a\sigma_i^b \sigma_j^a\sigma_j^b = &\sum_{ij}^{1,N} \sum_{ab}^{1,n} \sigma_i^a\sigma_i^b \sigma_j^a\sigma_j^b \\
    &- \sum_{i=1}^N \sum_{ab}^{1,n} (\sigma_i^a)^2(\sigma_i^b)^2 
\end{aligned}
\end{align*} 
we neglect the second sum, which leads to a $O(1/N)$ correction to the free energy density. Similarly in the following expression appearing in the ordered part of the partition function \eqref{A1-PartFunc1}
\begin{align*}
\begin{aligned}
    4! \sum_{i<j<k<l}^{1,N} &\sum_{a=1}^n\sigma_i^a\sigma_j^a\sigma_k^a\sigma_l^a = \sum_{ijkl}^{1,N}\sum_{a=1}^n\sigma_i^a\sigma_j^a\sigma_k^a\sigma_l^a \\
                            &- 6 \sum_{ijk}^{1,N}\sum_{a=1}^n (\cdots) - 4 \sum_{ij}^{1,N}\sum_{a=1}^n(\cdots) - \sum_{i=1}^N\sum_{a=1}^n(\cdots)
\end{aligned}
\end{align*}
we neglect all the sums apart from the first one, since they lead to corrections of order $O(1/N)$, $O(1/N^2)$ and $O(1/N^3)$ respectively to the free energy density. Therefore, the averaged expression of the replicated partition function at the leading order in $N$ reads
\begin{equation} \label{A1-PartFunc2}
  \begin{aligned}
      \overline{\Z^n} = \int \D \sigma &\exp\left[\frac{(\beta J)^2}{4 N} \sum_{ab}^{n,1} \left(\sum_{i=1}^N\sigma_i^a\sigma_i^b \right)^2\right] \times \\
                \times &\exp\left[\frac{\beta\epsilon}{N^3} \sum_{a=1}^n \left(\sum_{i=1}^N\sigma_i^a\right)^4\right].
  \end{aligned}  
\end{equation} 

At this point of the calculation the partition function depends only on two global parameters: the magnetization vector
\begin{align}
m_a = \frac{1}{N} \sum_{i=1}^N \sigma_i^a
\end{align}
and the overlap matrix with elements
\begin{align}
q_{ab} = \frac{1}{N} \sum_{i=1}^N \sigma_i^a\sigma_i^b.
\end{align}
By definition the matrix $\mathbb{Q}$ is symmetric and positive semidefinite and its diagonal elements are $q_{aa}=1$, due to the spherical constraint. These global parameters can be introduced in the computation of the partition function by exploiting the following identities
\begin{align} \label{A1-MagnetIntro}
1 &= \prod_{a=1}^n \int \de m_a\delta\left(Nm_a - \sum_i^N\sigma_i^a \right) \nonumber \\
&= \prod_{a=1}^n \int \de m_a \int_{-i \infty}^{+i\infty} \frac{N}{2\pi i} \de \rho_a e^{-\rho_a(Nm_a-\sum_i^N\sigma_i^a)}  \nonumber \\
&= \int \D m \int_{-i \infty}^{+i\infty}\D \rho~e^{-N \sum_a^n \rho_a m_a + \sum_a^n \sum_i^N \rho_a \sigma_i^a }
\end{align}
with
\begin{equation}
    \D m = \prod_{a=1}^n  \de m_a ~,~~~~~~ \D \rho = \prod_a^n \frac{N}{2\pi i} \de \rho_a
\end{equation}
and
\begin{align} \label{A1-OverlapIntro}
1 &= \prod_{a<b}^{1,n} \int \de q_{ab} \delta\left(Nq_{ab} - \sum_i^N \sigma_i^a\sigma_i^b \right)  \nonumber \\
&= \prod_{a<b}^{1,n} \int \de q_{ab} \int_{-i \infty}^{+i\infty} \frac{N}{2\pi i} \de \lambda_{ab} e^{-\lambda_{ab}(Nq_{ab} - \sum_i^N \sigma_i^a\sigma_i^b)} \nonumber \\
&= \int \D q  \int_{-i \infty}^{+i\infty}\D \lambda~e^{-\frac{N}{2}\sum_{a \neq b}^n \lambda_{ab}q_{ab}+ \frac{1}{2}\sum_{a \neq b}^n \sum_i^N \sigma_i^a \lambda_{ab} \sigma_i^b},
\end{align} 
with
\begin{equation}
    \D q = \prod_{a<b}^{1,n} \de q_{ab} ~,~~~~~~ \D \lambda = \prod_{a<b}^{1,n} \frac{N}{2\pi i} \de \lambda_{ab}.
\end{equation}
where the delta functions have been opened through a Laplace transformation and the conjugate variables of $m_a$ and $q_{ab}$ have been introduced, respectively $\rho_a$ and $\lambda_{ab}$. Due to the fact that $\mathbb{Q}$ is symmetric, the matrix $\Lambda$ (with elements $\lambda_{ab}$) is symmetric as well. Note that in the last line of Eq.~\eqref{A1-OverlapIntro} the sums in the exponential have been symmetrized . 

By exploiting a similar relation, the spherical constraint, which until now has been hidden in the definition \eqref{A1-Desigma}, can be opened as follows
\begin{align} \label{A1-Sph_constr-opening}
\prod_{a=1}^n &\delta\left(N-\sum_{i=1}^N (\sigma_i^a)^2\right) = \prod_a^n \int_{-i \infty}^{+i\infty} \frac{N}{4\pi i} \de \lambda_{aa}  \times \nonumber \\
&~~~~~~~~~~~~~~~~~~~~~~~~~~~~~\times e^{-\frac{\lambda_{aa}}{2} (N- \sum_i^N (\sigma_i^a)^2)}  \nonumber \\
&= \int_{-i \infty}^{+i\infty} \prod_a^n \frac{N}{4\pi i} \de \lambda_{aa} e^{-\frac{N}{2}\sum_a^n \lambda_{aa} q_{aa} + \frac{1}{2} \sum_a^n \sum_i^N \lambda_{aa}(\sigma_i^a)^2}.
\end{align}
Note that the previous expression for the spherical constraint perfectly matches with Eq.~\eqref{A1-OverlapIntro}. Hence, by neglecting constant prefactors and subleading contributions of order $O(\ln N/N)$, the expression of the replicated partition function reads
\begin{align}  \label{A1-PartFunc3}
\overline{\Z^n} &= \int \D \sigma \int \D q \int_{-i \infty}^{+i\infty}\D \lambda \int \D m \int_{-i \infty}^{+i\infty}\D \rho~\times  \nonumber \\
& \times \exp \Bigg[ \frac{(\beta J)^2 N}{4} \sum_{ab}^{1,n} q_{ab}^2 + \beta\epsilon N \sum_{a=1}^n m_a^4 - \frac{N}{2}\sum_{a b}^{1,n} \lambda_{ab}q_{ab}   \nonumber \\
&\quad+ \frac{1}{2}\sum_{ab}^{1,n} \sum_{i=1}^N \sigma_i^a \lambda_{ab} \sigma_i^b - N \sum_{a=1}^n \rho_a m_a + \sum_{i=1}^N \sum_{a=1}^n \rho_a \sigma_i^a \Bigg],
\end{align} 
where now
\begin{equation}
    \D\sigma = \prod_{i=1}^N\prod_{a=1}^n \de \sigma_i^a ~~~~~~ \D \lambda = \prod_{a \leq b}^{1,n} \frac{N}{2\pi i} \de \lambda_{ab}
\end{equation}

We notice that the spin dependent part of the action can be factorized with respect to the site indices, yielding the following expression 
\begin{equation*}
\exp\left[N\ln \int \prod_a^n d\sigma^a e^{\frac{1}{2}\sum_{ab}^n \sigma^a \lambda_{ab} \sigma^b + \sum_a^n \rho_a \sigma^a} \right].
\end{equation*}
The Gaussian integral in the spin variables can be easily performed 
\begin{align*}
\int \prod_a^n \de \sigma^a e^{\frac{1}{2}\sum_{ab}^n \sigma^a \lambda_{ab} \sigma^b + \sum_a^n \rho_a \sigma^a} = &\sqrt{\frac{(2 \pi)^n}  {\det(-\Lambda)}} \times  \\
&\times e^{-\frac{1}{2}\sum_{ab}^n\rho_a (\Lambda^{-1})_{ab} \rho_b}.
\end{align*}
Note that the previous integral is well defined only after shifting the integration of the $\Lambda$ elements in order for them to have a negative real part. This shift does not affect significantly the partition function. Eventually, by neglecting a constant, the partition function can be written as
\begin{align} \label{A1-PartFunc4}
\overline{\Z^n} &= \int \D q \int_{-i \infty}^{+i\infty}\D \lambda \int \D m \int_{-i \infty}^{+i\infty} \D \rho~e^{-N G[\mathbb{Q},\Lambda,{\bf m},\boldsymbol{\rho}]},
\end{align}
where the effective action $G$ has been defined as
\begin{align} \label{A1-Action}
G[\mathbb{Q},\Lambda,\bf{m},\boldsymbol{\rho}] &= - \frac{(\beta J)^2}{4} \sum_{ab}^{1,n} q_{ab}^2 - \beta\epsilon  \sum_{a=1}^n m_a^4 \nonumber \\
&\quad+ \frac{1}{2}\sum_{a b}^{1,n} \lambda_{ab}q_{ab}  +\sum_{a=1}^n \rho_a m_a \nonumber \\
&\quad+ \frac{1}{2}\sum_{ab}^{1,n} \rho_a (\Lambda^{-1})_{ab} \rho_b + \frac{1}{2}\ln\det(-\Lambda).
\end{align}

\subsubsection{The $\Lambda$ integration}
The integrals in Eq.~\eqref{A1-PartFunc4} can be performed through the saddle point method, which is why we only retained terms of order $O(1)$ in the large $N$ limit. The $\Lambda$-dependent part of eq.~\eqref{A1-Action} is 
\begin{equation} \label{A1-Action-lambda}
\begin{aligned}
    G_1[\Lambda] &= \frac{1}{2}\sum_{a b}^{1,n} \lambda_{ab}q_{ab}+ \frac{1}{2}\sum_{ab}^{1,n} \rho_a (\Lambda^{-1})_{ab} \rho_b \\
    &\quad + \frac{1}{2}\ln\det(-\Lambda).
\end{aligned}
\end{equation}
The following relation holds:

\begin{align}
\ln \det(-\Lambda- \boldsymbol{\rho} \otimes \boldsymbol{\rho}^T) &=  \ln\left[ (1+ \boldsymbol{\rho}^T \Lambda^{-1} \boldsymbol{\rho} )\det(-\Lambda) \right] \nonumber \\
&= \ln(1+ \boldsymbol{\rho}^T \Lambda^{-1} \boldsymbol{\rho} ) + \ln\det(-\Lambda) \nonumber \\
&= \ln\det(-\Lambda) + \boldsymbol{\rho}^T \Lambda^{-1} \boldsymbol{\rho} \nonumber\\
&\quad - \frac{1}{2}(\boldsymbol{\rho}^T \Lambda^{-1} \boldsymbol{\rho})^2 + O(n^3), \label{matrixrelation}
\end{align}

where we have used a general matrix relation for the determinant of
the sum of a matrix $\boldsymbol{A}$ and the external product of two
vectors $u$ and $v$, i.e.
$$\det(\boldsymbol{A}+uv^T)=(1+v^T \boldsymbol{A}^{-1} u)\det
\boldsymbol{A},$$ and then we have expanded the logarithm. We also
recall that $\boldsymbol{\rho} \otimes \boldsymbol{\rho}^T$ denotes a
matrix with elements $ (\boldsymbol{\rho} \otimes
\boldsymbol{\rho}^T)_{ab}=\rho_a\rho_b$. Thus, from the relation
\eqref{matrixrelation} one gets

\begin{equation*}
    \begin{aligned}
\ln\det(-\Lambda) + \boldsymbol{\rho}^T \Lambda^{-1} \boldsymbol{\rho} &= \ln \det(-\Lambda- \boldsymbol{\rho} \otimes \boldsymbol{\rho}^T) \\
&\quad + \frac{1}{2}(\boldsymbol{\rho}^T \Lambda^{-1} \boldsymbol{\rho})^2 + O(n^3).
\end{aligned}
\end{equation*}
By using the previous relation, eq.~\eqref{A1-Action-lambda} can be written as
\begin{equation}
  \begin{aligned}
    G_1[\Lambda] &= \frac{1}{2}\sum_{a b}^{1,n} \lambda_{ab}q_{ab} + \frac{1}{2}\ln\det(-\Lambda - \boldsymbol{\rho}\otimes \boldsymbol{\rho}^T) \\
    &\quad+ \frac{1}{4}\left( \sum_{ab}^{1,n} \rho_a (\Lambda^{-1})_{ab} \rho_b \right)^2 + O(n^3).
\end{aligned}  
\end{equation}
However, the relevant contribution to the free energy is of order $O(n)$, due to Eq.~\eqref{1-freeEnergy}. Thus the only part of $G_1[\Lambda]$ that we have to consider is
\begin{align}
G_1[\Lambda] = \frac{1}{2}\sum_{a b}^{1,n} \lambda_{ab}q_{ab} + \frac{1}{2}\ln\det(-\Lambda - \boldsymbol{\rho}\otimes \boldsymbol{\rho}^T).
\end{align}

The stationary point of $G_1[\Lambda]$ is, thus, given by
\begin{align}
\lambda_{ab}^* = -\rho_a\rho_b - (\mathbb{Q}^{-1})_{ab}
\end{align}
leading to
\begin{align}
G_1[\Lambda^*] = -\frac{1}{2}\sum_{ab}^{1,n} \rho_a q_{ab} \rho_b - \frac{1}{2} \ln\det \mathbb{Q},
\end{align}
where a constant has been neglected. Therefore, the complete effective action \eqref{A1-Action} reads as
\begin{equation}
    \begin{aligned}
    G[\mathbb{Q},\bf{m},\boldsymbol{\rho}] &= - \frac{(\beta J)^2}{4} \sum_{ab}^{1,n} q_{ab}^2 - \beta\epsilon  \sum_{a=1}^n m_a^4 +\sum_{a=1}^n \rho_a m_a \\
    &\quad -\frac{1}{2}\sum_{ab}^{1,n} \rho_a q_{ab} \rho_b - \frac{1}{2} \ln\det \mathbb{Q} .
\end{aligned}
\end{equation}

\subsubsection{The $\boldsymbol{\rho}$ integration}
Let us now perform the integration over $\boldsymbol{\rho}$. For this purpose the relevant part of the effective action is
\begin{align}
G_2[\boldsymbol{\rho}] = -\frac{1}{2}\sum_{ab}^{1,n} \rho_a q_{ab} \rho_b +\sum_{a=1}^n \rho_a m_a.
\end{align}
The stationary point of $G_2[\boldsymbol{\rho}]$ is given by the relation
\begin{align}
\rho_a^* = \sum_{b=1}^n (\mathbb{Q}^{-1})_{ab} m_b,
\end{align}
that leads to 
\begin{align}
G_2[\boldsymbol{\rho}^*] = \frac{1}{2}\sum_{ab}^{1,n} m_a (\mathbb{Q}^{-1})_{ab} m_b.
\end{align}
We have finally arrived to an expression of the effective action only in terms of the magnetization vector and the overlap matrix:
\begin{equation}
    \begin{aligned}
    G[\mathbb{Q},\bf{m}] &= - \frac{(\beta J)^2}{4} \sum_{ab}^{1,n} q_{ab}^2 - \beta\epsilon  \sum_{a=1}^n m_a^4 \\
           &\quad - \frac{1}{2} \ln\det \mathbb{Q} + \frac{1}{2}\sum_{ab}^{1,n} m_a (\mathbb{Q}^{-1})_{ab} m_b.
\end{aligned}
\end{equation}
We can now use the same trick used before to retain only the relevant terms for the saddle point. By exploiting the following relation
\begin{equation}
    \begin{aligned}
    \ln \det(\mathbb{Q}- {\bf m} \otimes {\bf m}^T) = & \ln\det \mathbb{Q} - {\bf m}^T \mathbb{Q}^{-1} {\bf m} \\
                 & - \frac{1}{2} ({\bf m}^T \mathbb{Q}^{-1} {\bf m})^2 + O(n^3),
\end{aligned}
\end{equation}
one gets

\begin{align}
& G[\mathbb{Q},{\bf m}] = - \frac{(\beta J)^2}{4} \sum_{ab}^n q_{ab}^2 - \beta\epsilon  \sum_a^n m_a^4  \nonumber \\
& - \frac{1}{2} \ln\det (\mathbb{Q}-{\bf m} \otimes {\bf m}^T) - \frac{1}{4} \left( \sum_{ab}^n m_a (\mathbb{Q}^{-1})_{ab} m_b \right)^2 + O(n^3).
\end{align}

However the only relevant contribution to the saddle point is given by terms of order $O(n)$, so that we have
\begin{equation}
    \begin{aligned}
G[\mathbb{Q},{\bf m}] = &- \frac{(\beta J)^2}{4} \sum_{ab}^n q_{ab}^2 - \beta\epsilon  \sum_a^n m_a^4 \\
&- \frac{1}{2} \ln\det (\mathbb{Q}-{\bf m} \otimes {\bf m}^T).
\end{aligned}
\end{equation}

\subsection{Replica symmetric free energy}
In order to solve the saddle point problem given by Eqs.~\eqref{1-SPeqs-a}, \eqref{1-SPeqs-b} we take the simplest possible ansatz on the structure of the matrix $\mathbb{Q}$, i.e.~the Replica Symmetric (RS) ansatz. We assume the overlap to be parametrized by only one variable
\begin{align}
q_{ab} = (1-q_0)\delta_{ab} + q_0 I_{ab},
\end{align} 
since the diagonal elements are $q_{aa}=1$ due to the spherical constraint. In the previous expression $I$ is a matrix whose elements are all ones. Moreover, we assume the magnetization vector to have all its components $m_a=m$. The energetic part of the action depending on the overlap can be then written as
\begin{align*}
\sum_{ab}^n q_{ab}^2 = n + n(n-1)q_0^2 = n (1-q_0^2) + O(n^2)
\end{align*}
and the entropic term as
\begin{align}
  & \ln\det (\mathbb{Q}-{\bf m}\otimes {\bf m}^T) = \nonumber \\
  & =\ln\left[ (1-q_0)^{n-1} (1-q_0+n(q_0-m^2)) \right] \nonumber \\
&= \ln\left[ (1-q_0)^n \left(1 + n \frac{q_0-m^2}{1-q_0} \right) \right]  \nonumber \\
&= n \ln(1-q_0) + \ln \left(1 + n \frac{q_0-m^2}{1-q_0} \right)  \nonumber \\
&= n \ln(1-q_0) + n \frac{q_0-m^2}{1-q_0} + O(n^2),
\end{align}
since the RS matrix $\mathbb{Q}-{\bf m}\otimes {\bf m}^T$ has only two kind of eigenvalues $\lambda_1=1-q_0$, with degeneracy $n-1$, and $\lambda_2=1-q_0+n(q_0-m^2)$. Hence, in the $n\rightarrow 0$ limit the RS effective action reads as
\begin{equation} 
    \begin{aligned} 
    \lim_{n\rightarrow 0} \frac{1}{n} G[q_0,m] &= - \frac{(\beta J)^2}{4} (1-q_0^2) - \beta\epsilon m^4 \\
&\quad- \frac{1}{2}\ln(1-q_0) - \frac{1}{2} \frac{q_0-m^2}{1-q_0}.
\end{aligned}
\end{equation}

The RS free energy is then given by
\begin{equation}
    f_{\rm RS} = \frac{1}{\beta} \lim_{n\rightarrow 0} \frac{1}{n} G[q_0,m].
\end{equation}

\bibliographystyle{unsrt}
\bibliography{biblio.bib}

\end{document}